\documentclass[11pt]{article}
\usepackage[a4paper,margin=1in]{geometry}
\usepackage[T1]{fontenc}
\usepackage[utf8]{inputenc}
\usepackage{microtype}
\usepackage{amsmath,amssymb}
\usepackage{graphicx}
\usepackage{booktabs}
\usepackage{multirow}
\usepackage{siunitx}
\usepackage{algorithm}
\usepackage{algpseudocode}
\usepackage{listings}
\usepackage{xcolor}
\usepackage{url}
\usepackage[hidelinks]{hyperref}
\usepackage[numbers,sort&compress]{natbib}
\usepackage[font=small,labelfont=bf,skip=4pt]{caption}
\usepackage{titlesec}
\usepackage{titling}
\usepackage[section,below]{placeins}
\usepackage{float}

\titleformat{\section}{\normalfont\Large\bfseries}{\thesection}{0.7em}{}
\titleformat{\subsection}{\normalfont\large\bfseries}{\thesubsection}{0.6em}{}

\pretitle{\begin{center}\LARGE\bfseries}
\posttitle{\par\end{center}\vspace{0.6em}}
\preauthor{\begin{center}\large}
\postauthor{\par\end{center}\vspace{0.0em}}
\predate{\begin{center}\normalsize}
\postdate{\par\end{center}\vspace{-0.2em}}

\title{CAFS: A Cache-Aware Frequency Sort for\\Low-Cardinality Integer Data on x86-64}
\author{Vasiliy S. Shlyk\\
{\normalsize Plekhanov Russian University of Economics}\\
{\normalsize Moscow, Russia}}
\date{May 2026}

\begin{document}
\maketitle

\renewcommand{\abstractname}{\normalsize Abstract}
\begin{abstract}\normalsize\noindent
Integer sorts in OLAP engines often run on columns whose cardinality $K$ is much smaller than the array length $N$. After a group-by stage the intermediate key column has $K$ bounded by the number of distinct group keys, and even a column-store scan typically operates on dictionary-encoded categorical fields where $K$ never exceeds a few thousand. A comparison sort on such a column still pays $\Theta(N \log N)$ comparisons, and a radix sort still pays $\Theta(N \cdot B/b)$ byte passes, irrespective of $K$. This paper describes CAFS, an integer sort that does exploit it on x86-64 with AVX2. The algorithm combines a SIMD bucket sized to one cache line, a Chao1 cardinality estimator over 1024 strided samples (kept in a heap-allocated 40\,KB open-addressing table), and an adaptive dispatcher backed by a spill safety guard. The hot loop is branchless and uses AVX2 cmpeq together with movemask and tzcnt to locate the matching lane. We benchmarked CAFS on a full-factorial grid of 58 array sizes $N$ from $10^3$ to $3 \cdot 10^7$ with dense $K$ schedules per $N$, producing 592770 timed runs against pdqsort, IPS4o, vqsort, ska\_sort, and std::sort. In the $K \ll N$ band the throughput is 1.7 to 3.1$\times$ that of pdqsort, 1.7 to 3.5$\times$ IPS4o, and 1.2 to 2.3$\times$ vqsort. The operational crossover against pdqsort is at $K \approx 1.3 \cdot 10^5$; against ska\_sort, $K \approx 8.14 \cdot 10^5$; against vqsort, $K \approx 6.7 \cdot 10^5$; and against IPS4o the curves only converge near $K = N$. Of the five baselines, only vqsort actually overtakes CAFS once the crossover is passed, which makes the vqsort threshold at $K \approx 6.7 \cdot 10^5$ the binding constraint on the operational range of CAFS.
\end{abstract}

\section{Introduction}

Sorting integer arrays serves as a core primitive in database management systems, OLAP engines, ETL pipelines, and column stores. Many of these workloads sort integer keys whose distinct-value count is small relative to the array length; common examples include status flags, event types, foreign keys into small dimension tables, and dictionary-encoded categorical attributes~\cite{boncz2005,stonebraker2005}. After a group-by stage, the intermediate key column has $K \ll N$ by construction, and a downstream sort runs on data with a heavy concentration of repeats. Similar patterns recur outside the database setting, for instance whenever a categorical column has been quantized down to a small alphabet.

Generic sorts ignore the property entirely. Comparison-based methods have an $\Omega(N \log N)$ lower bound on comparisons regardless of the input distribution~\cite{knuth1998,clrs2009}, and distribution sorts pay $\lceil B/b \rceil$ passes over key bytes (where $B$ is the key width and $b$ the radix digit width), so on 64-bit keys ska\_sort~\cite{skarupke} performs eight byte passes whether $K$ is 8 or $10^6$. On a $10^7$-element array with $K = 200$ a comparison sort issues about $2.3 \cdot 10^8$ comparisons, while the count-and-emit cost would be near $N + K \log K \approx 10^7$~\cite{knuth1998}, more than $20\times$ smaller. A single such sort completes in milliseconds, but at typical OLAP query rates of $10^3$--$10^4$ queries per second per node, the cumulative time spent inside the sort primitive becomes a non-trivial fraction of total CPU consumption.

Recent x86-64 cores make this gap exploitable in hardware. AVX2 gives 256-bit registers and instructions that compare four 64-bit keys in a single cycle, and BMI1 supplies tzcnt to extract the matching lane in two more cycles~\cite{intel2024}. Writing the hot loop with broadcast, cmpeq, movemask, and tzcnt also removes the conditional branches that would otherwise stall the pipeline by more than a dozen cycles per misprediction~\cite{intel2024,drepper2007}. The combined effect is a per-element cost of about one SIMD operation on a low-cardinality input, below the rate of any comparison or radix sort we are aware of.

We present CAFS (Cache-Aware Frequency Sort). CAFS uses a SIMD bucket on one cache line, a Chao1 cardinality estimator~\cite{chao1984} on a 1024-sample frequency set, and an adaptive dispatcher. The dispatcher routes to a tiny-count branchless path for $\hat K \le 8$, a hash-count main path for $8 < \hat K < N/2$, and a pdqsort~\cite{peters2021} fallback when $\hat K \cdot 2 > N$. A spill safety guard bounds the worst-case slowdown (Section~\ref{sec:cafs}). We evaluate CAFS empirically on a full-factorial $(N, K)$ grid against five baselines and report per-baseline crossover points $K^*$ above which CAFS no longer dominates. Our contributions are:

\begin{itemize}
\item A SIMD bucket layout fitting one cache line, with branchless update via AVX2 cmpeq, movemask, and tzcnt.
\item An adaptive dispatcher driven by a Chao1 cardinality estimator over 1024 strided samples, with a fixed-size FreqSet on the heap as the backing storage.
\item A three-branch hybrid (tiny-count for $\hat K \le 8$, main hash-count, pdqsort fallback for $\hat K \cdot 2 > N$) with a spill safety guard that caps the worst-case slowdown.
\item A header-only C++20 implementation for AVX2 on x86-64, with wrappers around pdqsort, IPS4o, vqsort, ska\_sort, and std::sort under a unified benchmark driver.
\item A full-factorial empirical study on a grid of 58 array sizes from $10^3$ to $3 \cdot 10^7$ with dense per-$N$ $K$ schedules, with explicit crossover points against each baseline.
\end{itemize}

Section~\ref{sec:bg} surveys the comparison, distribution, and SIMD sorting families together with the relevant cardinality estimators, and Section~\ref{sec:problem} fixes the formal model. The algorithm itself appears in Section~\ref{sec:cafs}, with pseudocode for the bucket update, the Chao1 estimator, the dispatcher, and the reconstruction step (the main hot loop is described in prose alongside its RLE behaviour). The implementation is described in Section~\ref{sec:impl}. Section~\ref{sec:eval} reports the per-baseline evaluation and the joint applicability map, Section~\ref{sec:disc} covers limitations and operating-point guidance, and Section~\ref{sec:concl} concludes.

\section{Background and Related Work}\label{sec:bg}

\subsection{Three families of integer sorts}

Integer sorts split naturally by their primitive decision step~\cite{knuth1998,clrs2009}. Comparison sorts build the order through $x < y$ and trace paths in a decision tree, which forces $\Omega(N \log N)$ comparisons in the worst and average cases for any input distribution. Distribution sorts read each key directly and place it in the output at an index computed from its value, at a typical cost of $O(N \cdot B/b)$, where $B$ is the key bit width and $b$ the radix digit width. Hash-count sorts reduce sorting to frequency counting followed by ordered emission of (key, count) pairs. Their total cost decomposes as $T(N, K) = T_{\text{count}}(N) + T_{\text{sort}}(K) + T_{\text{emit}}(N)$, which collapses to $O(N)$ with a small constant whenever $K \ll N$~\cite{knuth1998}.

Table~\ref{tab:family_comparison} arranges the main competitors along three axes: asymptotic cost, dependence on $K$, and use of SIMD. Algorithms that ignore $K$ pay their full asymptotic cost on every input. Algorithms that exploit $K$ win at $K \ll N$, but the classical implementations do not vectorize the hot loop and lose to comparison and radix sorts on $K \approx N$ because of the high constant in the count step. CAFS exploits $K$ and also vectorizes the hot loop, with a deliberate fallback to a strong comparison sort on high-$K$ inputs.

\begin{table}[!htb]
\centering
\caption{Integer sort families compared on three axes.}
\label{tab:family_comparison}
\begin{tabular}{l l c c l}
\toprule
Algorithm & Cost & Uses $K$? & SIMD & Extra memory \\
\midrule
std::sort (introsort)~\cite{musser1997} & $O(N \log N)$ & no & no & $O(\log N)$ \\
pdqsort~\cite{peters2021}                & $O(N \log N)$ & no & partial & $O(\log N)$ \\
IPS4o (sequential)~\cite{axtmann2017}    & $O(N \log N)$ & no & no & $O(\log N)$ (in-place) \\
ska\_sort (LSD radix)~\cite{skarupke}    & $O(N \cdot B/b)$ & no & no & $O(N + 2^b)$ \\
vqsort (Highway)~\cite{wassenberg2022}   & $O(N \log N)$ & no & AVX2/AVX-512 & $O(\log N)$ \\
generic hash-count                       & $O(N + K \log K)$ & yes & no  & $O(K)$ \\
CAFS (this paper)                        & $O(N + K \log K)$ & yes & AVX2 & $O(K)$ \\
\bottomrule
\end{tabular}
\end{table}

\subsection{Comparison sorts}

The C++ standard library ships introsort~\cite{musser1997} as \texttt{std::sort}, which combines quicksort with a heapsort safeguard at recursion depth $2 \log_2 N$ and an insertion-sort cutoff under 16 elements. Pattern-defeating quicksort (pdqsort)~\cite{peters2021} replaces the classical Lomuto partition with a branchless Hoare-style partition that eliminates the conditional store; pdqsort additionally detects already-sorted, reverse-sorted, and almost-sorted prefixes and falls back to a different algorithm on adversarial inputs. On random inputs the branchless partition removes the dominant cost (a mispredicted comparison stalls a modern x86 pipeline for more than a dozen cycles~\cite{intel2024,drepper2007}) and yields a 1.5 to 3 times speedup over introsort. IPS4o~\cite{axtmann2017} replaces the binary partition with an in-place $k$-way super-scalar samplesort whose classifier is a balanced binary tree built from a sampled splitter set; the predictable memory access pattern lets it outrun pdqsort by 1.2 to 1.5 times on random 64-bit integers in single-threaded mode. The branchless partitions in both pdqsort and IPS4o derive from BlockQuicksort~\cite{edelkamp2016}. None of the three algorithms exploits low $K$.

\subsection{Distribution sorts}

Counting sort needs $O(N + R)$ memory and time, which is infeasible for $R = 2^{64}$. Radix sorts decompose the key into $b$-bit digits and apply counting sort per digit~\cite{knuth1998}. The least-significant-digit (LSD) variant scans from low to high in $\lceil B/b \rceil$ passes. The most-significant-digit (MSD) variant sorts the high digits first and recurses on the resulting buckets, which permits early exit on short buckets and admits an in-place implementation. The ska\_sort implementation~\cite{skarupke} uses an in-place American-flag MSD radix with type-specialized inner loops, signed-integer handling, and an insertion-sort cleanup on short buckets; on 64-bit keys this amounts to eight byte passes of bucket rearrangement that touch every element on each pass, with the per-pass move count of order $N$ and the constant depending on the bucket distribution. Subsequent work in this family includes the AVX-friendly radix sorts of Satish et al.~\cite{satish2010} for CPUs and GPUs with vectorized digit extraction, and the SIMD-plus-cache-friendly merge sort of Inoue and Taura~\cite{inoue2015} for column-store workloads. The cost of every algorithm in this family scales with the key width $B$ and is independent of $K$.

\subsection{SIMD sorts}

SIMD-aware sorts push the hot loop onto vector lanes without adapting to $K$. Among earlier proposals, Bramas~\cite{bramas2017} built a hybrid quicksort on AVX-512 with sorting networks at the leaves, and Lemire et al.~\cite{lemire2014} showed that SIMD compression and decompression beat scalar baselines by an order of magnitude on integer streams, which already shows that vector lanes do useful work in sort-related kernels. The current strongest entry is vqsort~\cite{wassenberg2022}, which drives a recursive partition through AVX2 or AVX-512 sorting networks on YMM/ZMM registers and merges sorted halves with masked stores. Its hot loop runs near 0.5 to 1 cycle per element against 4 to 5 cycles for scalar pdqsort, and on 64-bit random inputs it outpaces prior comparison and radix sorts by roughly a factor of two.

\subsection{Cardinality estimation}

The Chao1 estimator~\cite{chao1984} infers the unseen part of a population from singletons and doubletons in a sample. The original form is $\hat K = u + f_1^2 / (2 f_2)$ and the standard bias-corrected variant is $\hat K = u + f_1 (f_1 - 1) / (2 (f_2 + 1))$. Both are undefined or unstable when $f_2 = 0$, so the implementation in this paper uses the simpler smoothed variant
\[
\hat K_{\text{Chao1}} = u + \frac{f_1^2}{2 \, (f_2 + 1)},
\]
where $u$ is the count of distinct values seen in the sample, $f_1$ is the count of values seen exactly once, and $f_2$ is the count of values seen exactly twice. The $f_2 + 1$ keeps the formula well-defined when $f_2 = 0$, and we use $f_1^2$ in the numerator rather than the canonical $f_1(f_1-1)$ for implementation simplicity. The two expressions differ by $f_1 / (2(f_2+1))$, which can reach the order of $f_1/2$ when $f_2$ is small, but in that saturated regime the dispatcher already routes to the high-entropy fallback through the $\hat K \cdot 2 > N$ guard, so the choice does not affect dispatching behaviour. HyperLogLog~\cite{flajolet2007} reaches relative standard error around $1.04/\sqrt m$ with $m$ registers, which is asymptotically tight but expensive at the fixed memory footprint we target (under 4 KB for the estimator). CAFS uses Chao1 because the algorithm needs only an order-of-magnitude estimate of $K$ within microseconds, which a 1024-sample Chao1 on a 4096-slot table delivers.

\subsection{Cache and SIMD context}

Two earlier results inform the cache and SIMD analysis below. Frigo et al.~\cite{frigo1999} formalized cache-oblivious algorithms that match the optimal cache behaviour up to constants. Drepper~\cite{drepper2007} catalogued the relevant x86 memory-hierarchy effects, namely branch-misprediction cost and the L1, L2, L3, and DRAM access latencies that govern the constants in our hot-loop analysis. Cycle counts for individual AVX2 and BMI1 instructions used in the bucket update are taken from the Intel Optimization Reference Manual~\cite{intel2024}.

\section{Problem Setting}\label{sec:problem}

\subsection{Inputs and goal}

Given an array $X = (x_1, \ldots, x_N)$ of 64-bit unsigned integers from a value set $V \subseteq \mathbb{Z}_{\ge 0}$, the task is to compute a permutation $\pi$ with $x_{\pi(i)} \le x_{\pi(i+1)}$ for $1 \le i < N$. Stability is not required: equal integer keys are indistinguishable. We write $K = |\{x_i : 1 \le i \le N\}|$ for the cardinality, $2 \le K \le N$, and restrict attention to the regime $K \ll N$ that is typical of analytical workloads~\cite{boncz2005,stonebraker2005}.

\subsection{Information-theoretic gap}

The Shannon entropy per key under a uniform distribution over $K$ values is $H = \log_2 K$ bits, so a 64-bit integer array with $K$ unique values uses only $H$ of its 64 bits per element. The gap $\Delta = 64 - H$ is information capacity unused by the data but still loaded into registers and cache on every access. Neither comparison nor distribution sorts can recover this gap. A comparison sort issues $\log_2 N$ comparisons per element and a radix sort visits $\lceil B/b \rceil$ digits per key whatever $H$ may be. With a hash function of adequate spread and a table large enough to keep buckets sparse, a hash-count scheme drops the per-element cost to $O(H)$ comparisons, scaling with the information content rather than with the key width or the array length.

\subsection{Cost model}

A hash-count sort decomposes into three components:
\begin{equation}
T(N, K) = T_{\text{count}}(N) + T_{\text{sort}}(K) + T_{\text{emit}}(N).
\label{eq:tnk}
\end{equation}
For $K \ll N$ the middle term $T_{\text{sort}}(K) = O(K \log K)$ is negligible, and under a good count primitive the total cost approaches $T(N, K) \approx 2 N$. For $K \approx N$ the middle term rises to $O(N \log N)$, the count overhead remains, and the comparison family regains the asymptotic advantage. Equation~\eqref{eq:tnk} thus covers the dominance and fallback regimes of CAFS with a single expression.

\subsection{Entropy bin}

We aggregate measurements by entropy bin
\begin{equation}
H_{\text{bin}} = \lfloor \log_2 K \rfloor.
\label{eq:hbin}
\end{equation}
Bin 1 covers $K \in [2, 4)$, bin 2 covers $[4, 8)$, up to bin 24 for $[2^{23}, 2^{24})$. Bin index is the information density of the input in bits per key. Aggregation by $H_{\text{bin}}$ turns the $(N, K)$ scatter into a heatmap and absorbs the per-bin noise from the exact $K$ value within the bin.

\subsection{Speedup, win rate, and crossover}

For a baseline $X$ at point $(N, K)$, the speedup is
\begin{equation}
\text{speedup}_X(N, K) = \frac{t_X(N, K)}{t_{\text{CAFS}}(N, K)}.
\label{eq:speedup}
\end{equation}
A value above 1 means CAFS wins. Inside an entropy bin we report the arithmetic mean (avg), minimum (min), maximum (max), and the win rate, defined as the fraction of points in the bin with $\text{speedup} > 1$. The operational crossover against baseline $X$ is the largest $K$ (at $N > 10^6$) for which CAFS wins on at least 50\% of grid points in the corresponding entropy bin. We use the win rate rather than the bin mean to avoid sensitivity to outliers with extreme speedup. The Pearson correlation between $\log N$ and speedup inside the dominance zone indicates whether the lead grows, shrinks, or stays flat with $N$.

\subsection{Methodology}

The grid covers 58 values of $N$ from $10^3$ to $3 \cdot 10^7$, stepped in four piecewise-linear bands: 1000 to 50000 (step 2000), 50000 to 1 million (step 50000), 1 million to 10 million (step 1 million), and 10 million to 30 million (step 5 million). For each $N$ the cardinality $K$ is enumerated from 2 to $N$ along a piecewise-coarsening schedule: step 1 below $K=200$, step 10 from 200 to 15000, step 500 from 15000 to $10^5$, and $\max(5000, K/10)$ above that, which gives a dense sweep at small $K$ and a roughly geometric sweep once $K$ grows past the L1 working set. Every algorithm is run twice on a fresh copy of the input and we keep the minimum of the two runs to suppress scheduler noise. Correctness is checked against a reference array sorted by std::sort and any point with a mismatch is dropped from aggregation, although CAFS produced no such failure on the grid. Each $(N, K)$ point is tested against five baselines plus CAFS, and together with the rerun rows for low-entropy bins and the per-bin book-keeping rows the resulting CSV holds 592770 measurement rows.

\section{The CAFS Algorithm}\label{sec:cafs}

\subsection{Overview and pipeline}

The algorithmic skeleton is the count-emit pair: count the frequency of each unique value, then emit (key, count) pairs in increasing key order. The hot loop never compares two input values with $<$; the only equality check it performs is between an input value and a bucket slot. The count step always costs $O(N)$ at one SIMD operation per element, so when $K \ll N$ the whole algorithm runs near that rate, whereas at $K \approx N$ the trailing pair sort dominates and a dedicated comparison engine wins.

\begin{figure}[!htb]
\centering
\includegraphics[width=0.92\textwidth]{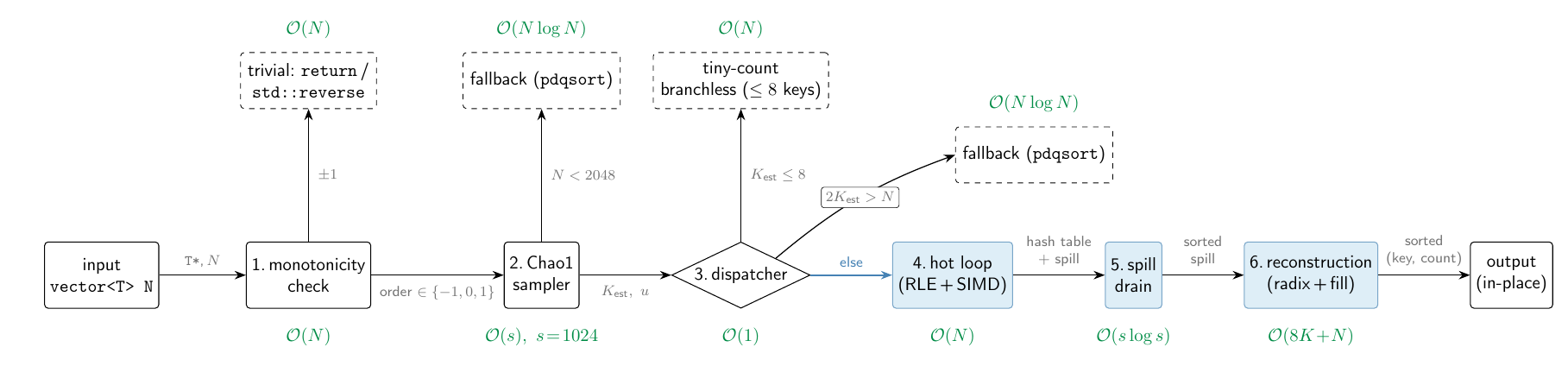}
\caption{CAFS pipeline.}
\label{fig:pipeline}
\end{figure}

The pipeline runs in seven stages, drawn in Figure~\ref{fig:pipeline}. The monotonicity check returns early on already-sorted input through a fast scan that exits at the first inversion (on random input that happens at the second or third element). The sampler then reads 1024 strided samples and feeds a heap-allocated open-addressing frequency set FreqSet of 4096 slots (about 40\,KB for 64-bit keys), from which the Chao1 estimator computes $\hat K$, $u$, $f_1$, and $f_2$. On those values the dispatcher selects a branch (Algorithm~\ref{alg:dispatcher}), the selected branch iterates over the input and updates buckets through the SIMD primitive of Section~\ref{sec:bucket}, reconstruction folds the buckets and the spill into a dense pair array, and a final emit step expands those pairs into the output via std::fill\_n.

The monotonicity check pays $O(N)$ comparisons with early exit. Sampling and Chao1 add $O(s)$ at $s = 1024$, under 100 microseconds on a modern x86 core, and the dispatcher itself is a single comparison. The tiny-count branch pays $O(N)$ with eight branchless adds per element, while the main hot loop pays roughly one SIMD operation per element. Spill drain costs $O(|\text{spill}| \log |\text{spill}|)$ when the spill is non-empty and nothing otherwise. The trailing pair sort runs std::sort if $K' < 256$ and an 8-pass LSD byte radix otherwise.

\subsection{The SIMD bucket}\label{sec:bucket}

\begin{figure}[!htb]
\centering
\includegraphics[width=0.7\textwidth]{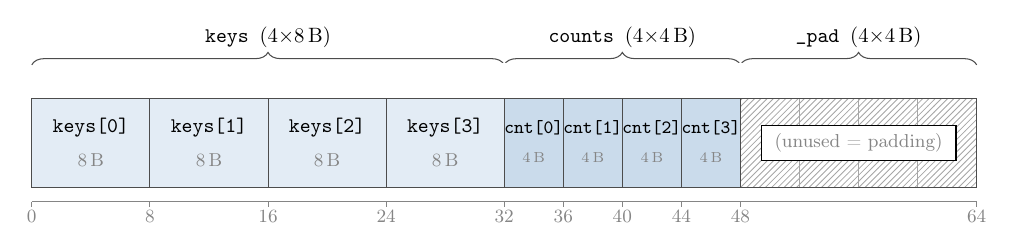}
\caption{Layout of \texttt{Bucket<uint64\_t>}: 64 bytes aligned to a cache line.}
\label{fig:bucket}
\end{figure}

The bucket is a fixed-capacity record aligned to a 64-byte cache line. For \texttt{uint64\_t} the bucket holds four 64-bit keys (32 B), four 32-bit counters (16 B), and 16 B of padding. For \texttt{int32\_t} it holds eight keys and eight counters, exposing one AVX2 instruction across all eight lanes. Figure~\ref{fig:bucket} shows the layout.

\begin{figure}[!htb]
\centering
\includegraphics[width=0.85\textwidth]{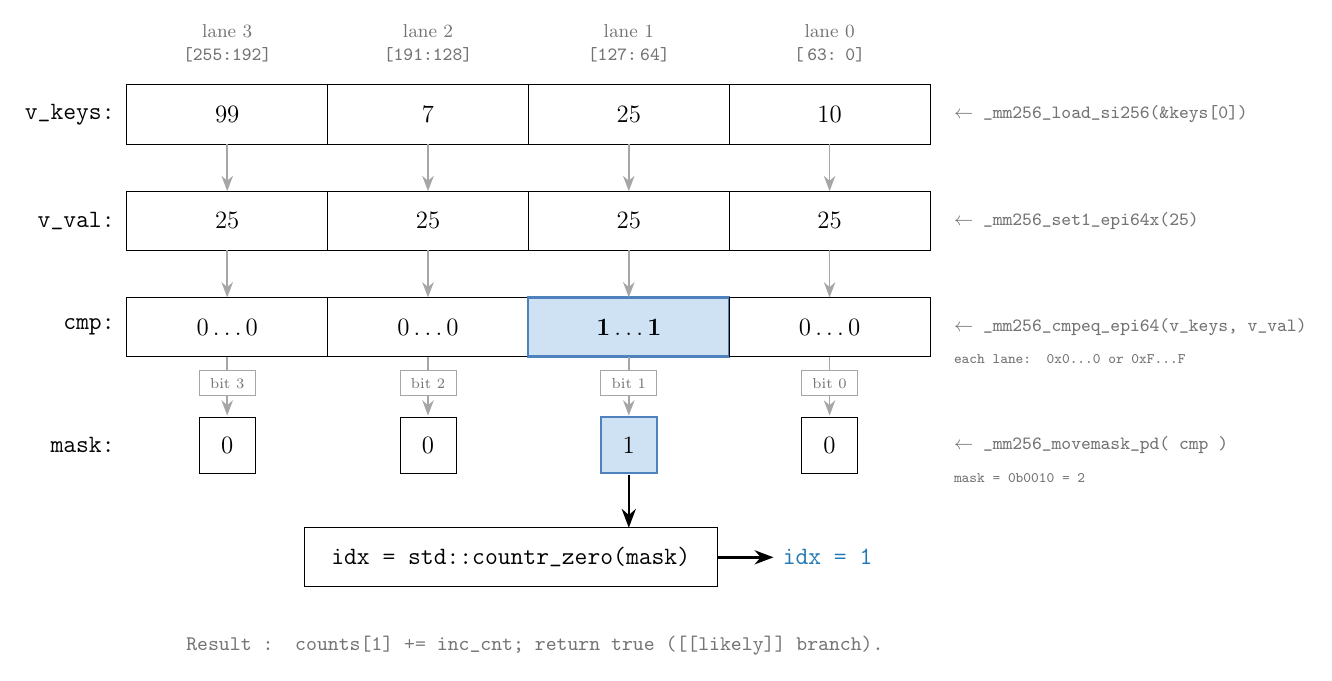}
\caption{Branchless bucket update through AVX2 instructions.}
\label{fig:hotpath}
\end{figure}

The \texttt{update(val, inc)} operation takes a value and an increment count. On a hit it adds $\textit{inc}$ to the matching counter, on a miss it claims the first empty slot (the one whose counter is zero), and on a fully occupied bucket it returns false so the value is sent to the spill. The 64-bit specialization compiles to five instructions: \texttt{\_mm256\_load\_si256} pulls four keys into a YMM register, \texttt{\_mm256\_set1\_epi64x} broadcasts the query value across the four lanes, \texttt{\_mm256\_cmpeq\_epi64} compares them in parallel, \texttt{\_mm256\_movemask\_pd} packs the result into a 4-bit mask, and \texttt{std::countr\_zero} (BMI1 tzcnt) returns the index of the matched lane. The data flow is drawn in Figure~\ref{fig:hotpath} and the operation is stated in Algorithm~\ref{alg:bucket}. With per-instruction latencies on Alder Lake~\cite{intel2024} of 1 cycle for \texttt{vmovdqa} from L1, 1 for \texttt{vpbroadcastq}, 1 for \texttt{vpcmpeqq}, 2 for \texttt{vmovmskpd}, 3 for \texttt{tzcnt}, and 1 for the counter add, the critical-path dependency on a hit (\texttt{vpcmpeqq} → \texttt{vmovmskpd} → \texttt{tzcnt} → counter add) sums to 7 cycles, which out-of-order overlap across successive iterations amortizes to roughly 4 to 5 cycles per element.

\begin{algorithm}[!htb]
\caption{Branchless bucket update for \texttt{uint64\_t} on AVX2.}
\label{alg:bucket}
\begin{algorithmic}[1]
\Function{Update}{bucket, val, inc}
  \State $\textit{keys} \gets \texttt{\_mm256\_load\_si256}(\text{bucket.keys})$
  \State $\textit{q} \gets \texttt{\_mm256\_set1\_epi64x}(\text{val})$
  \State $\textit{eq} \gets \texttt{\_mm256\_cmpeq\_epi64}(\textit{keys}, \textit{q})$
  \State $\textit{m} \gets \texttt{\_mm256\_movemask\_pd}(\text{cast}(\textit{eq}))$
  \If{$\textit{m} \ne 0$}
    \State $j \gets \texttt{tzcnt}(\textit{m})$
    \State $\text{bucket.counts}[j] \mathrel{+}= \textit{inc}$ \Comment{hit path: 4--5 cycles}
    \State \Return $\textit{true}$
  \EndIf
  \For{$j = 0$ to $\text{CAP} - 1$} \Comment{cold path: claim empty slot}
    \If{$\text{bucket.counts}[j] = 0$}
      \State $\text{bucket.keys}[j] \gets \text{val}$
      \State $\text{bucket.counts}[j] \gets \textit{inc}$
      \State \Return $\textit{true}$
    \EndIf
  \EndFor
  \State \Return $\textit{false}$ \Comment{full bucket, value goes to spill}
\EndFunction
\end{algorithmic}
\end{algorithm}

The bucket does not resolve in-bucket collisions. When a bucket fills, the next miss spills to a side buffer. A Poisson model with capacity 4 and mean fill 0.5 puts the probability of overflow at $P(X \ge 5) \approx 1.7 \cdot 10^{-4}$, so spills stay rare across the input distributions in our grid (Section~\ref{sec:eval}). The hash function is the multiplicative hash with the inverse golden ratio:
\begin{equation}
h(x) = \left\lfloor \frac{(x \cdot \varphi^{-1}) \bmod 2^{64}}{2^{64 - \log_2 M}} \right\rfloor, \quad \varphi^{-1} = \texttt{0x9E3779B97F4A7C15},
\label{eq:hash}
\end{equation}
which compiles into one multiplication and one shift~\cite{knuth1998}. The choice of hash matters because low-cardinality columns in practice exhibit arithmetic structure (sequential foreign keys, codes from a contiguous range $\{0, \ldots, K-1\}$, or fixed-step quantized features), and the naive $h(x) = x \bmod M$ collides catastrophically on such inputs because a linear progression maps to every $M$-th bucket. Multiplicative hashing distributes a linear progression across the index space pseudo-uniformly and cuts the collision rate by about three orders of magnitude at the same $M$.

\subsection{Cardinality estimation}

Before the main path runs, CAFS samples the input at stride $\max(1, \lfloor N/s \rfloor)$ with $s = 1024$, then drops the samples into a heap-allocated open-addressing table FreqSet of 4096 slots (40\,KB for 64-bit keys: a 32\,KB key array plus an 8\,KB counter array). Three statistics fall out: $u$ is the count of occupied slots, $f_1$ is the count of slots with frequency 1, and $f_2$ is the count of slots with frequency 2. Algorithm~\ref{alg:chao1} states the estimator.

\begin{algorithm}[!htb]
\caption{Chao1 cardinality estimator on a fixed-size FreqSet.}
\label{alg:chao1}
\begin{algorithmic}[1]
\Function{Chao1}{$X$, $N$, $s = 1024$}
  \State allocate FreqSet on the heap (4096 slots, all empty)
  \State $\textit{stride} \gets \max(1, \lfloor N / s \rfloor)$
  \For{$i = 0, \textit{stride}, 2 \textit{stride}, \ldots$ until $\min(N, s \cdot \textit{stride})$}
    \State linear-probe insert $X[i]$ into FreqSet, increment its counter
  \EndFor
  \State $u \gets$ count of occupied slots
  \State $f_1 \gets$ count of slots with counter $= 1$
  \State $f_2 \gets$ count of slots with counter $= 2$
  \If{$u = s$}
    \State \Return $(N, u, f_1, f_2)$ \Comment{saturated; high-entropy input}
  \EndIf
  \State \Return $\bigl(u + f_1^2 / (2(f_2 + 1)),\ u,\ f_1,\ f_2\bigr)$
\EndFunction
\end{algorithmic}
\end{algorithm}

\begin{figure}[!htb]
\centering
\includegraphics[width=0.8\textwidth]{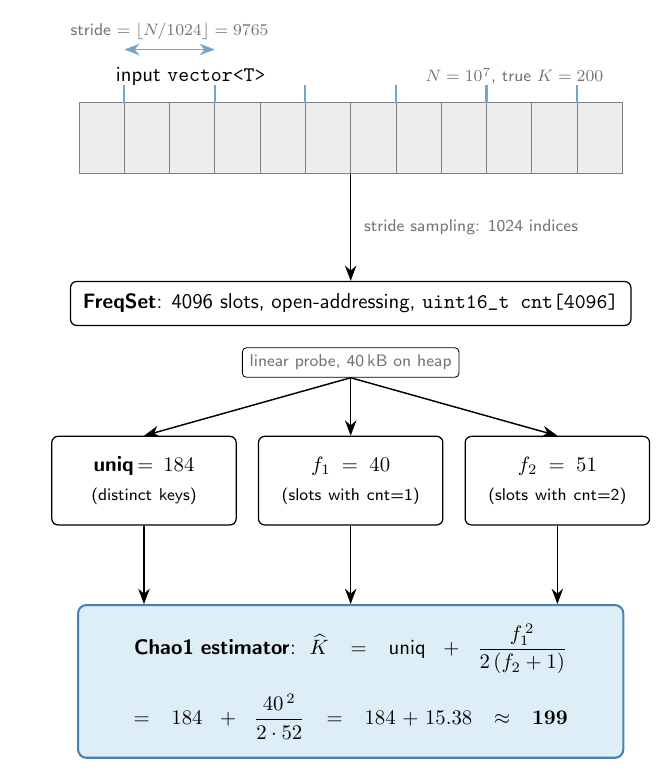}
\caption{The Chao1 estimator: sample, FreqSet, frequency spectrum, $\hat K$.}
\label{fig:chao1}
\end{figure}

The Chao1 estimator~\cite{chao1984} reads
\begin{equation}
\hat K_{\text{Chao1}} = u + \frac{f_1^2}{2 \, (f_2 + 1)}.
\label{eq:chao1}
\end{equation}
The correction term $f_1^2 / (2(f_2 + 1))$ infers the unseen part of the population from the spectrum of rare values in the sample. The Laplace smoothing $f_2 + 1$ avoids division by zero when no doubletons appear; the asymptotic effect is negligible at $f_2$ in the tens or higher. When $u = s$ the estimator saturates to $N$, which routes the dispatcher to the high-entropy path. The full estimation costs 1024 hash computes, 1024 probes, and 4096 counter reads, under 100 microseconds on a modern x86 core. Figure~\ref{fig:chao1} shows the data flow.

\subsection{Adaptive dispatcher and hot loop}

\begin{figure}[!htb]
\centering
\includegraphics[width=0.85\textwidth]{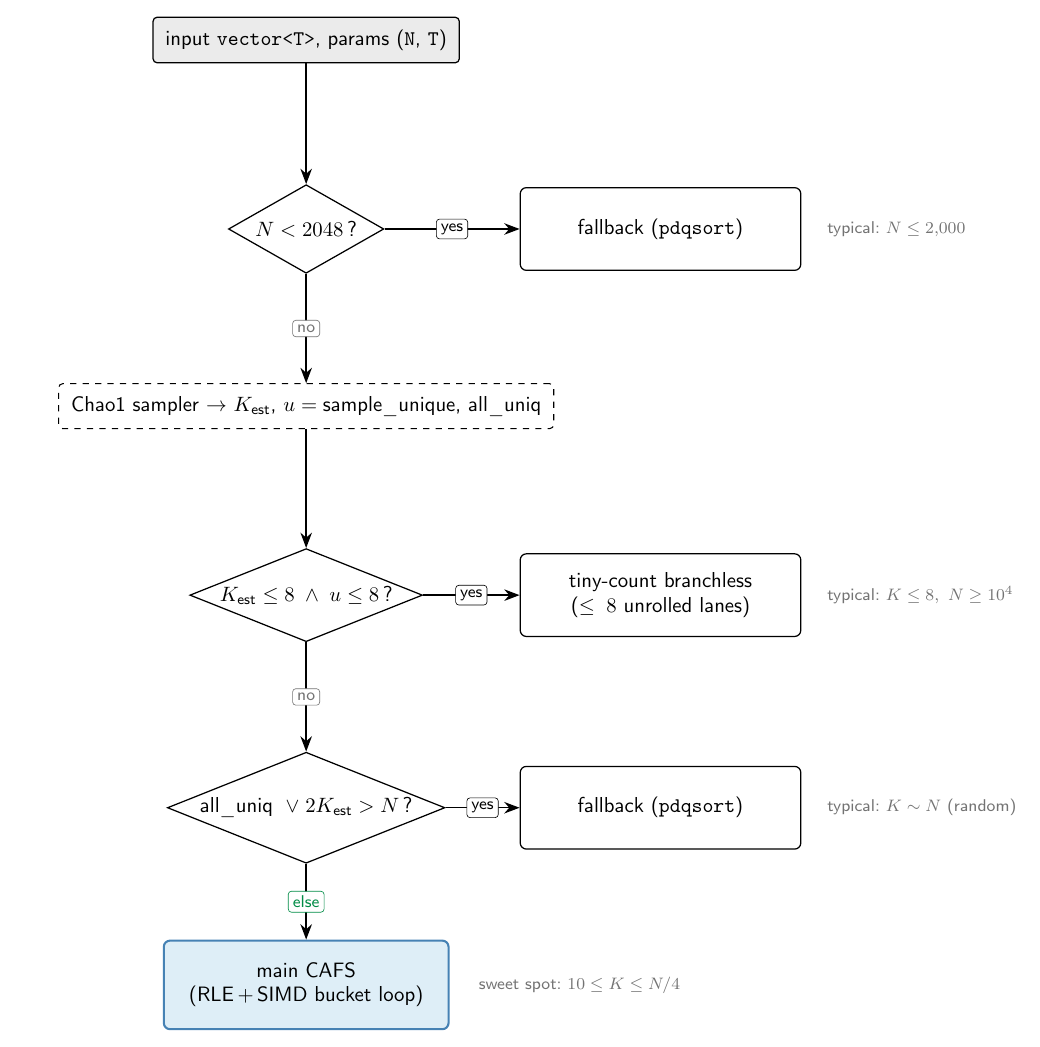}
\caption{Dispatcher decision tree.}
\label{fig:dispatcher}
\end{figure}

The dispatcher reads $\hat K$, $u$, and $N$ and routes to one of three content branches plus a trivial bypass. Table~\ref{tab:dispatcher} lists the conditions; Algorithm~\ref{alg:dispatcher} encodes the same logic. Figure~\ref{fig:dispatcher} draws the decision tree.

\begin{table}[!htb]
\centering
\caption{Dispatcher decisions and the selected branch.}
\label{tab:dispatcher}
\begin{tabular}{lll}
\toprule
Condition & Branch & Rationale \\
\midrule
$N < 2048$ & fallback (pdqsort) & sampling does not amortize \\
$\hat K \le 8 \wedge u \le 8$ & tiny-count branchless & 8 stack counters beat the SIMD bucket \\
$\hat K \cdot 2 > N$ & fallback (pdqsort) & high entropy, hash scheme not viable \\
otherwise & main CAFS & $8 < \hat K < N/2$ \\
\bottomrule
\end{tabular}
\end{table}

\begin{algorithm}[!htb]
\caption{Adaptive dispatcher.}
\label{alg:dispatcher}
\begin{algorithmic}[1]
\Function{Dispatch}{$X$, $N$}
  \If{\Call{IsMonotone}{$X$, $N$}}
    \State \Return early on already-sorted input \Comment{trivial bypass}
  \EndIf
  \If{$N < 2048$}
    \State \Return pdqsort($X$, $N$) \Comment{pre-pass too expensive at small $N$}
  \EndIf
  \State $(\hat K, u, f_1, f_2) \gets$ \Call{Chao1}{$X$, $N$}
  \If{$\hat K \le 8 \wedge u \le 8$}
    \State \Return \Call{TinyCount}{$X$, $N$, sample keys}
  \EndIf
  \If{$\hat K \cdot 2 > N$}
    \State \Return pdqsort($X$, $N$) \Comment{high-entropy fallback}
  \EndIf
  \State \Return main hot loop ($X$, $N$, $\hat K$) \Comment{described in prose below}
\EndFunction
\end{algorithmic}
\end{algorithm}

\FloatBarrier

The dispatcher itself is a single comparison after the sampler. The pre-pass pays at most $3 \cdot 10^5$ cycles in total. When the condition $\hat K \le 8 \wedge u \le 8$ fires, the algorithm extracts up to eight keys from the sample and runs eight independent counters with branchless updates. Each comparison $(v == k_j)$ compiles into a cmp/setcc pair without a conditional jump. The eight add instructions are register-independent and execute concurrently on the out-of-order back end, which yields about 1 to 2 cycles per element on random 64-bit input. After the pass, the branch checks $\sum c_j = N$. On success it emits the output via std::fill\_n. On failure (Chao1 underestimated and the input has a value outside the eight sampled keys) control falls into the main path.

\begin{figure}[!htb]
\centering
\includegraphics[width=0.8\textwidth]{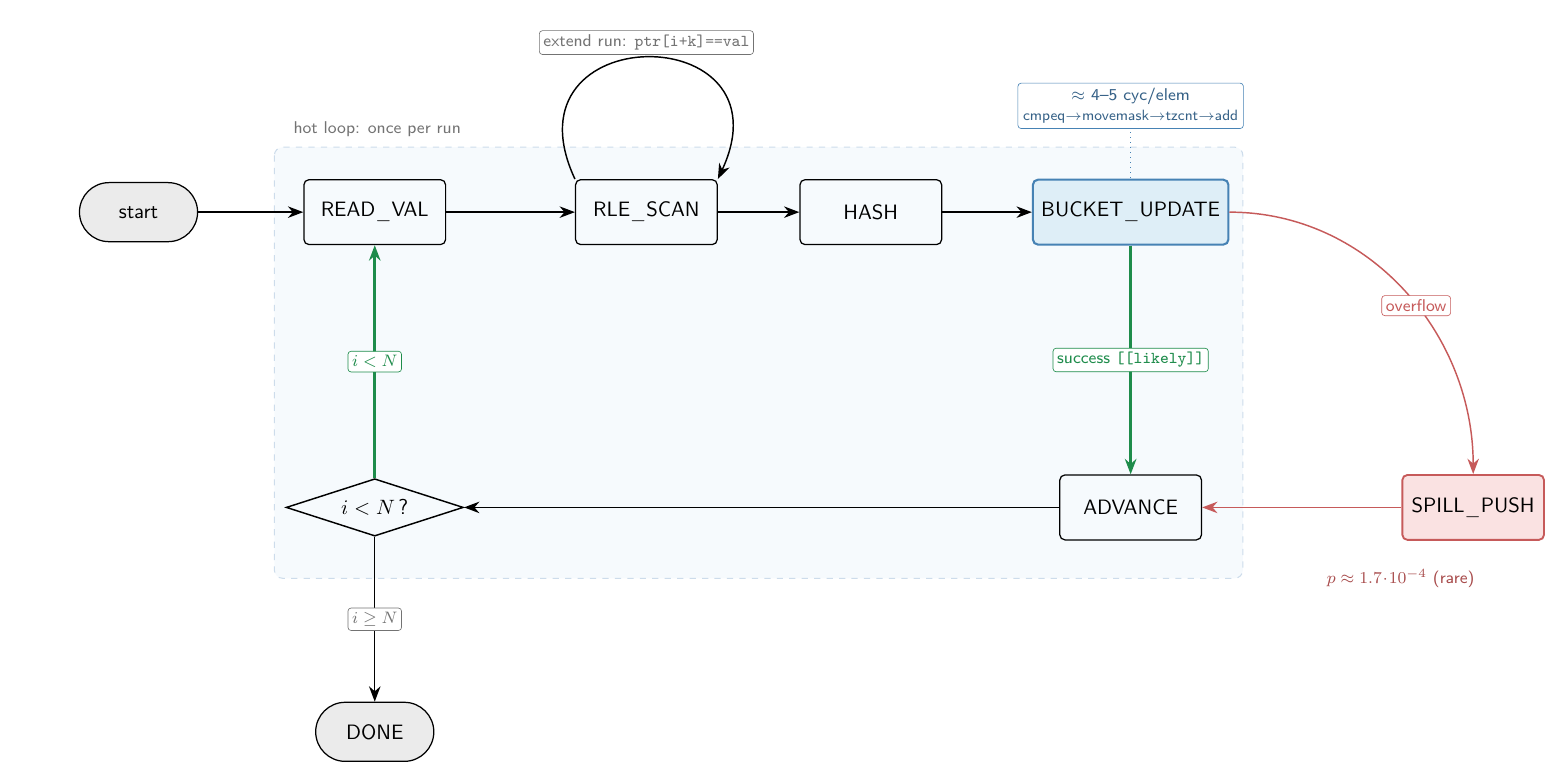}
\caption{Main hot loop state: pointer ptr, index $i$, run\_len, active bucket.}
\label{fig:hotloop}
\end{figure}

The main path also runs a run-length-encoding (RLE) pass over consecutive equal elements: before each \texttt{update} call CAFS extends the run and replaces $r$ separate calls with a single \texttt{update(val, run\_len)}. The benefit depends on input shape. On fully random input with $K \gg 1$ runs degenerate to length 1 and RLE adds one comparison per element without reducing the \texttt{update} count, whereas on sorted-by-group inputs the average run is $N/K$ and the call count collapses from $N$ to $K$ (with already-sorted input as the degenerate end of the same curve). Figure~\ref{fig:hotloop} shows the loop state.

\subsection{Reconstruction and complexity}

\begin{figure}[!htb]
\centering
\includegraphics[width=0.85\textwidth]{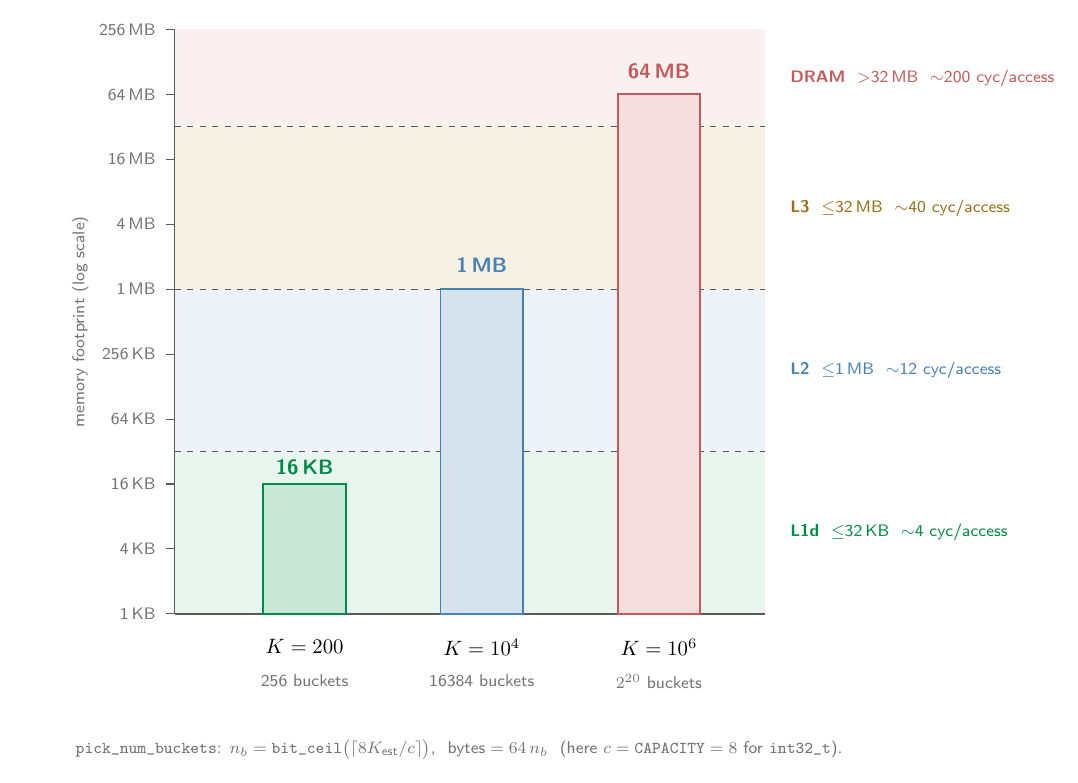}
\caption{Hash table size as a function of $\hat K$, illustrative. The horizontal markers are drawn at generic x86 L1d / L2 / L3 sizes; the test platform (Table~\ref{tab:setup}) has a 48 KB L1d, which shifts the L1d marker relative to the figure. The exact bucket counts for our $\hat K = 200, 4\cdot 10^3, 10^4, 10^6$ examples are given in the text below; figure values are sketched at illustrative resolution and may differ by a power of two from the formula in the text.}
\label{fig:Msize}
\end{figure}

The hash table size adapts to the estimate:
\begin{equation}
M = \text{bit\_ceil}(8 \cdot \hat K / \text{cap}),
\label{eq:Msize}
\end{equation}
where cap is 4 for 64-bit keys and 8 for 32-bit keys. The target slot load factor is $1/8$, which gives an expected full-bucket rate near $0.1\%$ under a Poisson allocation. The size is clamped between $M \ge 8$ and $M \le \text{bit\_ceil}(N / \text{cap})$. Figure~\ref{fig:Msize} maps $M$ to the cache hierarchy of the test platform.

At $\hat K = 200$, $M = 512$ buckets give a 32 KB table that fits comfortably inside the 48 KB L1d of the test platform (Table~\ref{tab:setup}). At $\hat K = 4 \cdot 10^3$, $M = 8192$ gives a 512 KB table that fits in the 1.25 MB L2. At $\hat K = 10^4$, $M = 32768$ gives a 2 MB table, which exceeds the L2 and lives in L3. At $\hat K = 10^6$, $M = 2^{21}$ is 128 MB and goes to DRAM, but at that scale the dispatcher condition $\hat K \cdot 2 > N$ usually fires and CAFS routes to the fallback before the hash table is even allocated.

\begin{figure}[!htb]
\centering
\includegraphics[width=0.85\textwidth]{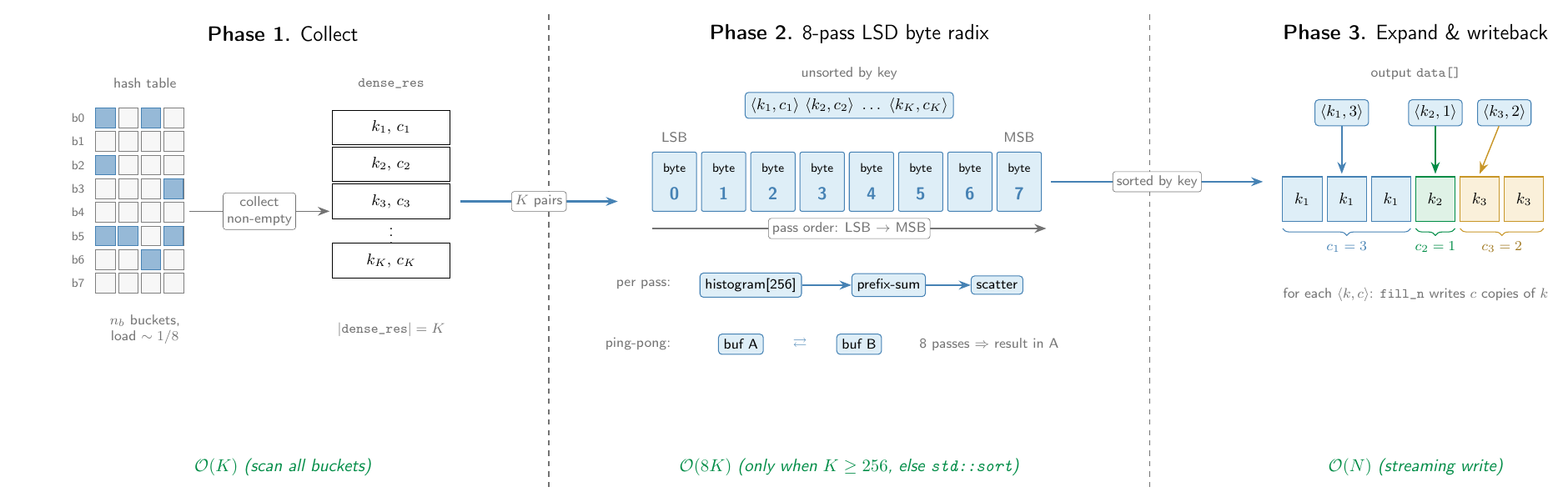}
\caption{Reconstruction: buckets, spill, dense\_res, radix tail.}
\label{fig:reconstruct}
\end{figure}

Reconstruction after the hot loop has three steps. (a) If $|\text{spill}| > N/2$ the result is dropped and CAFS restarts with the pdqsort fallback. This guard caps the worst case at roughly twice pdqsort time, and the absolute worst case observed in our grid was about 5 times. (b) Otherwise the spill is sorted with std::sort, equal neighbors are folded into (key, count) pairs, and the resulting pairs are appended to the dense bucket pairs. (c) The combined dense array of size $K' \le \hat K + |\text{spill folded}|$ is then sorted by key. The procedure is stated in Algorithm~\ref{alg:reconstruct} and the data path is drawn in Figure~\ref{fig:reconstruct}.

\begin{algorithm}[!htb]
\caption{Reconstruction with spill safety guard.}
\label{alg:reconstruct}
\begin{algorithmic}[1]
\Function{Reconstruct}{buckets, spill, $N$, out}
  \If{$|\text{spill}| > N / 2$}
    \State \Return pdqsort($X$, $N$) \Comment{safety guard fires}
  \EndIf
  \State sort spill with std::sort
  \State fold equal neighbors of the spill into (key, count) pairs $\to$ \texttt{spill\_pairs}
  \State collect non-empty bucket slots into pairs $\to$ \texttt{dense\_pairs}
  \State \texttt{dense\_pairs} $\gets$ \texttt{dense\_pairs} $\cup$ \texttt{spill\_pairs}
  \State $K' \gets |\texttt{dense\_pairs}|$
  \If{$K' < 256$}
    \State sort \texttt{dense\_pairs} by key with std::sort
  \Else
    \State sort \texttt{dense\_pairs} by key with 8-pass LSD byte radix
  \EndIf
  \State $p \gets \text{out}$
  \For{each $(k, c)$ in \texttt{dense\_pairs}}
    \State \texttt{std::fill\_n}($p$, $c$, $k$); $p \gets p + c$
  \EndFor
\EndFunction
\end{algorithmic}
\end{algorithm}

The pair sort runs std::sort with a key comparator when $K' < 256$ and an 8-pass LSD byte radix when $K' \ge 256$. The expansion uses std::fill\_n(out, count, key) per pair, which compiles to rep stosq or a SIMD store burst with throughput close to L2/L3 bandwidth (30 to 50 GB/s).

\begin{figure}[!htb]
\centering
\includegraphics[width=0.85\textwidth]{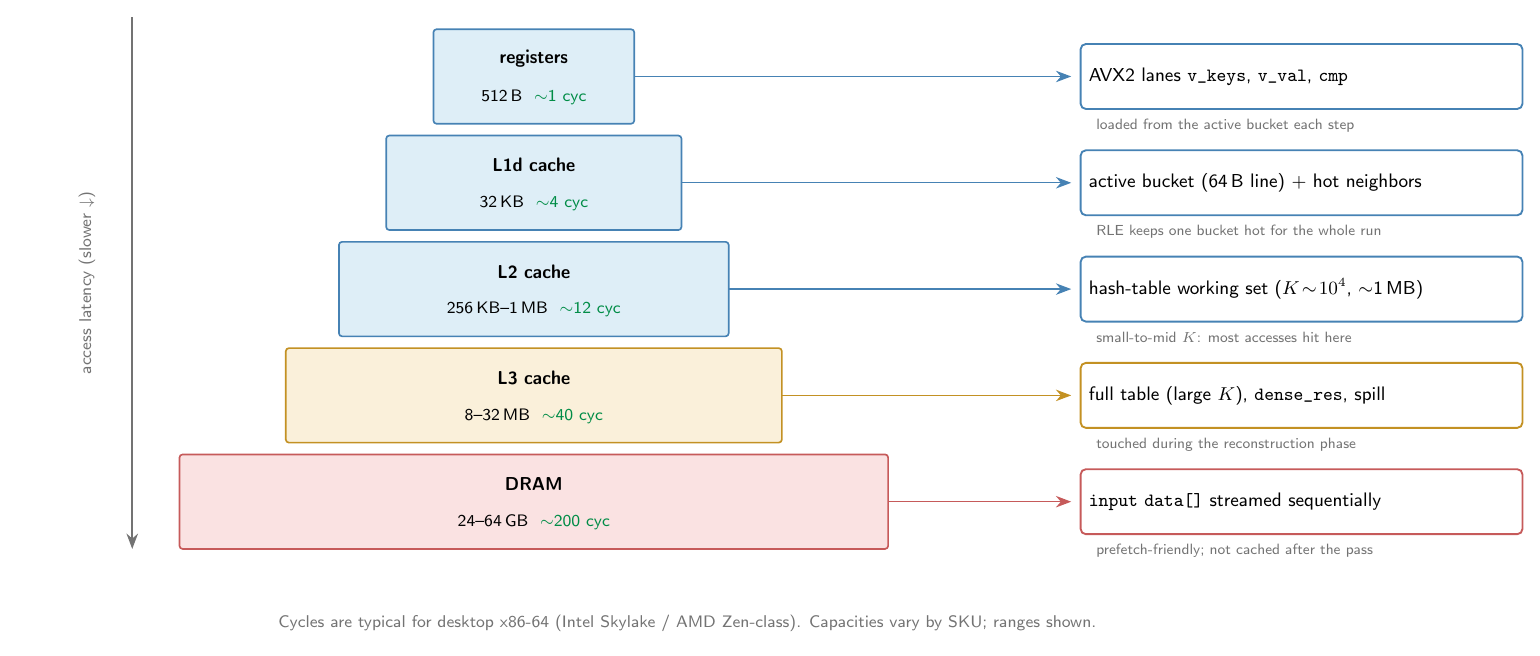}
\caption{CAFS working set across the x86-64 cache hierarchy.}
\label{fig:hierarchy}
\end{figure}

The branch-wise time complexities are:
\begin{align}
T_{\text{tiny}}(N) &= O(N), \label{eq:Ttiny}\\
T_{\text{main}}(N, K) &= O(N) + O(K \log K), \label{eq:Tmain}\\
T_{\text{main+radix}}(N, K) &= O(N) + O(8 K), \label{eq:Tmainradix}\\
T_{\text{fallback}}(N) &= O(s) + O(N \log N). \label{eq:Tfb}
\end{align}
Total memory is dominated by the bucket table at $64M$ bytes (one 64-byte cache line per bucket). The expected spill size on the input distributions in our grid is $O(N \cdot 1.7 \cdot 10^{-4})$, with a hard cap of $N/2$ enforced by the guard. The dense pair array dense\_res holds at most $K + |\text{spill folded}|$ entries, and the radix buffer consumes a further $K' \cdot \text{sizeof(Pair)}$ bytes.

\section{Implementation}\label{sec:impl}

CAFS ships as a header-only C++20 library. It pulls only \texttt{<algorithm>}, \texttt{<bit>}, \texttt{<cstdint>}, \texttt{<cstring>}, \texttt{<vector>}, \texttt{<utility>}, and \texttt{<immintrin.h>}. The C++20 surface used is \texttt{std::bit\_ceil}, \texttt{std::bit\_width}, and \texttt{std::countr\_zero}. The target ISA is x86-64 with AVX2; SIMD specializations exist for \texttt{int32\_t}, \texttt{int64\_t}, and \texttt{uint64\_t}. Other integral types fall back to a scalar four-key linear scan with no loss of correctness. Builds were verified on g++ 13.2 (MinGW-w64) on Windows 11 with \texttt{-O3 -mavx2 -mbmi -std=c++20 -DNDEBUG}.

All baselines enter the benchmark through thin wrappers in \texttt{competitors.hpp}. std::sort is taken from libstdc++. Both pdqsort~\cite{peterspdq} and ska\_sort~\cite{skarupke} are single-header libraries and are included directly. IPS4o~\cite{ips4ocode} is also distributed header-only, but its internal scheduler instantiates TBB types even on the sequential entry point; the wrapper supplies a stub header with empty TBB-type declarations to satisfy the includes, after which only the sequential entry point is called. vqsort is part of Google Highway~\cite{highway} and is built via \texttt{add\_subdirectory} in CMake, linking statically against the \texttt{hwy} and \texttt{hwy\_contrib} targets. The i5-12400F does not expose AVX-512, so the Highway runtime dispatch lands on the AVX2 implementation, which is the same ISA target as CAFS.

\paragraph{Artifact availability.}
The implementation, the benchmark driver, the baseline wrappers, and the raw measurement CSVs are released under the MIT license at \url{https://github.com/kexibq-official/cafs-lib}. The repository also contains build scripts for libhwy and the vqsort object files, together with the Python notebooks used to produce the figures and tables. The source layout is summarized in Table~\ref{tab:files}.

\begin{table}[!htb]
\centering
\caption{Source layout of the implementation.}
\label{tab:files}
\begin{tabular}{ll}
\toprule
File & Contents \\
\midrule
cafs.hpp & primitives: Bucket<T> with SIMD specializations, fast\_hash, prototype v1 \\
cafs2.hpp & adaptive dispatcher, FreqSet, Chao1 estimator, tiny-count, radix tail \\
competitors.hpp & wrappers for pdqsort, ska\_sort, IPS4o, vqsort \\
datagen.hpp & low-entropy input generator with controlled $K$ \\
main\_bigdata.cpp & benchmark driver: $(N, K)$ grid, correctness check, CSV export \\
\bottomrule
\end{tabular}
\end{table}

\section{Evaluation}\label{sec:eval}

\subsection{Goals, hypotheses, methodology}

The evaluation answers three questions. First, what is the bin-mean speedup of CAFS against each baseline at $K \ll N$, and is it large enough to be useful? Second, where, in terms of $K$, does each baseline overtake CAFS at $N > 10^6$? Third, is the lead at fixed $K$ stable across $N$, or does it grow or shrink as the array gets bigger?

For each point $(N, K)$ a palette of $K$ values is built as the arithmetic progression $a + b \cdot i$ for $i = 0, \ldots, K-1$, where $a$ and an odd step $b$ are derived from the seed $42 + N + K$ through a SplitMix-style mixer, and the array of length $N$ is then filled by uniform sampling from this palette, with each algorithm receiving its own fresh copy. Runs are timed with \texttt{std::chrono::high\_resolution\_clock} (reported as floating-point milliseconds), each algorithm is executed twice on every point regardless of $N$, and the minimum of the two replicates enters the result table. Correctness is verified by sorting a separate copy with std::sort and comparing element-wise, and CAFS produced no failures across the grid. The order of algorithms within a point is fixed at (CAFS, std::sort, pdqsort, ska\_sort, IPS4o, vqsort) and we did not randomize it; the methodological consequences are discussed in Section~\ref{sec:bench-platform}.

We benchmark CAFS against five single-threaded baselines: pdqsort and IPS4o (comparison), vqsort (SIMD), ska\_sort (radix), and std::sort (the C++ standard library default). Each $(N, K)$ point produces one speedup value per baseline, and we aggregate by entropy bin $H_{\text{bin}} = \lfloor \log_2 K \rfloor$.

\subsection{Bench platform}\label{sec:bench-platform}

The bench is single-threaded, background processes were minimized between runs, and the P-state lock was not enforced, so thermal throttling adds about $\pm 5\%$ noise across long runs (Table~\ref{tab:setup}). The fixed algorithm order within each $(N, K)$ point (Section~\ref{sec:eval}) introduces an additional systematic bias whose sign is the opposite of what one might assume: CAFS runs first on a cold cache and the baselines that follow benefit from a working set that is already paged in. The effect is at most a few percent of run time on the largest inputs and below measurement noise once the working set exceeds L3. We do not claim this bias cancels in the speedup ratio; it is a methodological limitation of the present study.

\begin{table}[!htb]
\centering
\caption{Bench platform and toolchain.}
\label{tab:setup}
\begin{tabular}{ll}
\toprule
Parameter & Value \\
\midrule
CPU & Intel Core i5-12400F, 6 P-cores, base 2.5 GHz, boost 4.4 GHz \\
ISA extensions & x86-64, AVX2, BMI1, BMI2 \\
L1d cache & 48 KB per core, 12-way associative \\
L2 cache & 1.25 MB per core, 10-way associative \\
L3 cache & 18 MB shared, 12-way associative \\
RAM & 32 GB DDR4-3200 \\
OS & Windows 11 \\
Compiler & g++ 13.2 (MinGW-w64) \\
Build flags & -O3 -mavx2 -mbmi -std=c++20 -DNDEBUG \\
\bottomrule
\end{tabular}
\end{table}

\FloatBarrier

\subsection{CAFS vs.\ pdqsort}

Across bins 1 to 16 the bin-mean speedup is between 1.7 and 3.1, with the win rate between 67\% and 96\%. The win rate stays above 90\% all the way through bin 11 (the safe zone) and only drops to 67\% by bin 16, where the bin mean is still above 1.6. Bin 17 is the transition (mean 1.29, win rate 56\%), and parity sets in from bin 18 onward. The minimum entries between 0.20 and 0.39 mark the points where the safety guard fired and CAFS finished at about half the speed of pdqsort because of the sampling overhead; no point exceeded a $5\times$ slowdown. The last dominance point at $N > 10^6$ is at $K \approx 1.72 \cdot 10^7$ ($H \approx 24$ bits), and the operational crossover (the largest $K$ where the win rate still clears 50\%) corresponds to $H \approx 17$ bits ($K \approx 1.3 \cdot 10^5$). Inside the dominance zone the Pearson correlation between $\log N$ and speedup is $r = 0.232$, so the lead trends mildly upward with $N$.

\begin{figure}[H]
\centering
\includegraphics[width=0.82\textwidth]{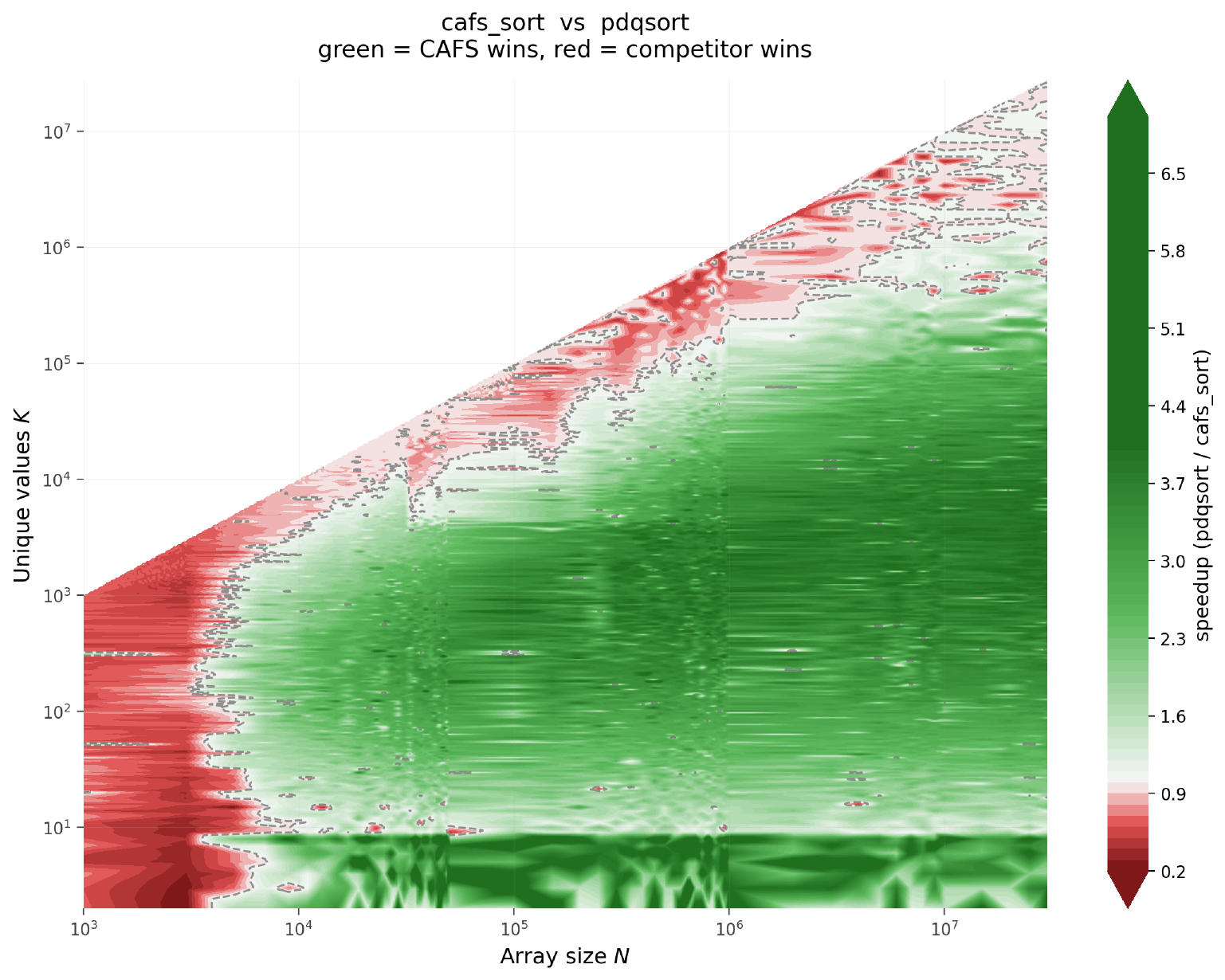}
\caption{Heatmap of CAFS speedup over pdqsort across the $(N, K)$ grid.}
\label{fig:vs_pdqsort}
\end{figure}

\begin{table}[H]
\centering
\caption{Speedup of CAFS over pdqsort, by entropy bin.}
\label{tab:vs_pdqsort}
\begin{tabular}{c c r r r r}
\toprule
$H_{\text{bin}}$ & $K$ range & avg & min & max & win-rate \\
\midrule
1 & 2 to 4 & 2.50 & 0.20 & 7.38 & 94.8\% \\
2 & 4 to 8 & 3.10 & 0.29 & 6.49 & 94.8\% \\
3 & 8 to 16 & 1.72 & 0.14 & 6.36 & 88.4\% \\
4 & 16 to 32 & 1.86 & 0.25 & 3.72 & 92.9\% \\
5 & 32 to 64 & 2.30 & 0.26 & 4.23 & 94.6\% \\
6 & 64 to 128 & 2.63 & 0.20 & 4.92 & 93.6\% \\
7 & 128 to 256 & 2.86 & 0.25 & 8.09 & 94.9\% \\
8 & 256 to 512 & 3.00 & 0.27 & 5.16 & 94.3\% \\
9 & 512 to 1024 & 3.07 & 0.24 & 6.84 & 94.4\% \\
10 & 1024 to 2048 & 3.01 & 0.24 & 5.93 & 95.6\% \\
11 & 2048 to 4096 & 2.78 & 0.30 & 7.31 & 93.4\% \\
12 & 4096 to 8192 & 2.35 & 0.21 & 9.57 & 87.2\% \\
13 & 8192 to 16384 & 2.12 & 0.24 & 6.30 & 71.5\% \\
14 & 16384 to 32768 & 1.92 & 0.34 & 4.35 & 67.1\% \\
15 & 32768 to 65536 & 1.87 & 0.29 & 4.23 & 81.5\% \\
16 & 65536 to 131072 & 1.66 & 0.33 & 3.54 & 77.4\% \\
17 & 131072 to 262144 & 1.29 & 0.34 & 3.14 & 55.9\% \\
18 & 262144 to 524288 & 1.01 & 0.39 & 2.29 & 41.7\% \\
$\ge 19$ & $\ge$ 524288 & $\le$ 1.00 & 0.36 & 1.71 & $\le$ 49.6\% \\
\bottomrule
\end{tabular}
\end{table}

\FloatBarrier

\subsection{CAFS vs.\ IPS4o}

The bin-mean speedup in bins 1 to 11 ($K$ from 2 to 4096) is between 1.97 and 3.52 at win rates above 97\%, with a maximum of 3.52 in bin 9 ($K = 512$ to 1024). At that $K$ the hash table fits in L1d, RLE injection stays inactive on random input, and the sampling cost is small relative to the count step. Through bins 12 to 16 (up to $K = 131072$) the mean falls to between 1.7 and 2.5 with win rates between 79\% and 96\% as the table moves out of L1d through L2 and into L3 while IPS4o keeps its prefetch-friendly partition pattern. Bin 17 is the transition (mean 1.33, win rate 58\%), and from bin 18 onward the bin means are near 1.00. Unlike the pdqsort and vqsort pairings, there is no sharp crossover against IPS4o; the lead decreases smoothly to parity, with the last dominance point at $N > 10^6$ corresponding to $K \approx 1.72 \cdot 10^7$ ($H \approx 24$ bits), close to $K = N$. The Pearson correlation between $\log N$ and speedup inside the dominance zone is $r = -0.189$, a slight tilt downward as $N$ grows.

Hypothesis H2 thus holds against a strong comparison engine, not only against introsort. The 1.97 to 3.52 range across the safe band exceeds the 2$\times$ threshold of H1 in most of bins 1 to 11.

\begin{figure}[H]
\centering
\includegraphics[width=0.82\textwidth]{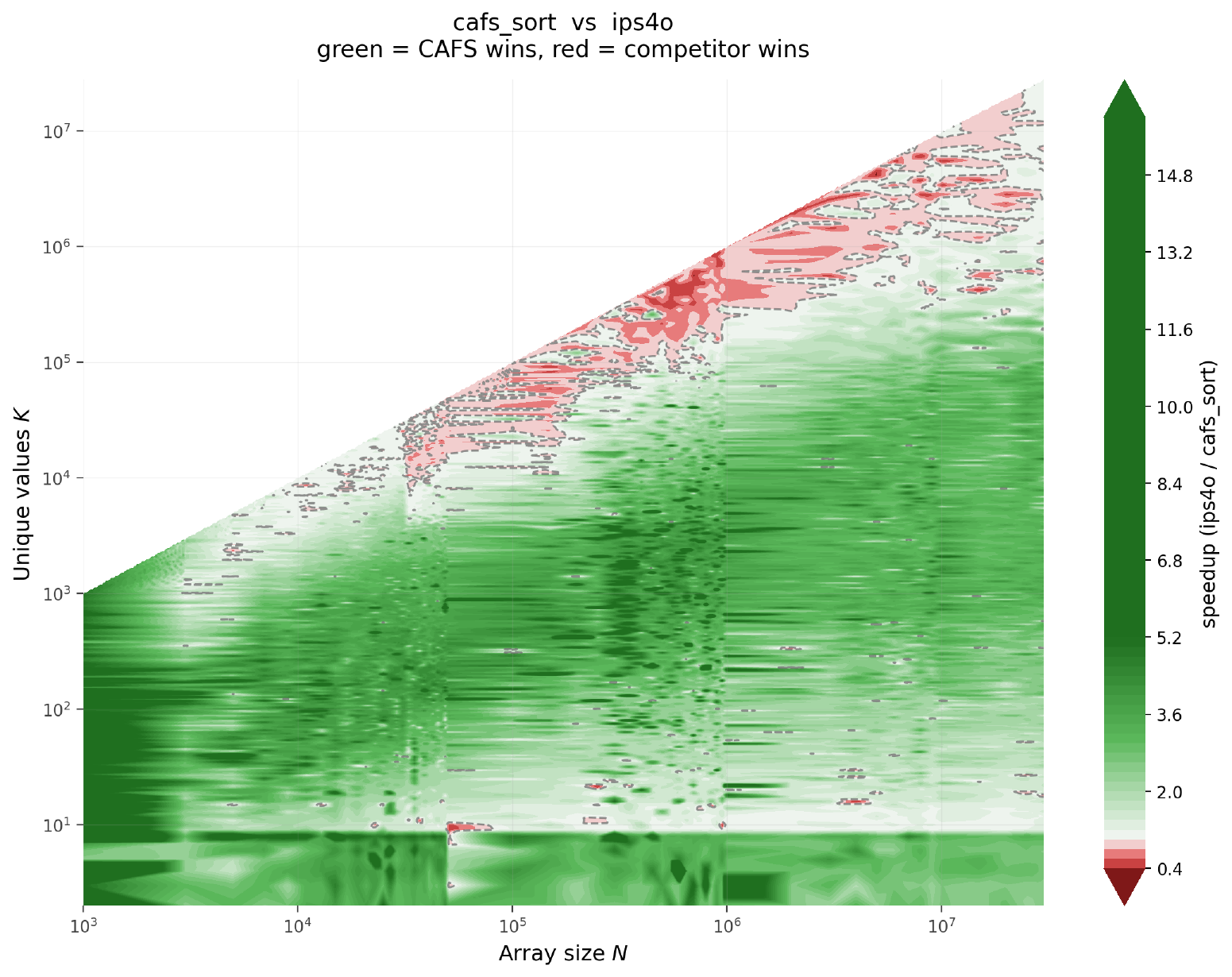}
\caption{Heatmap of CAFS speedup over IPS4o across the $(N, K)$ grid.}
\label{fig:vs_ips4o}
\end{figure}

\begin{table}[H]
\centering
\caption{Speedup of CAFS over IPS4o, by entropy bin.}
\label{tab:vs_ips4o}
\begin{tabular}{c c r r r r}
\toprule
$H_{\text{bin}}$ & $K$ range & avg & min & max & win-rate \\
\midrule
1 & 2 to 4 & 3.24 & 0.65 & 11.64 & 99.1\% \\
2 & 4 to 8 & 3.50 & 0.70 & 17.50 & 99.6\% \\
3 & 8 to 16 & 1.97 & 0.24 & 15.85 & 97.0\% \\
4 & 16 to 32 & 2.09 & 0.22 & 13.36 & 97.4\% \\
5 & 32 to 64 & 2.57 & 0.19 & 10.95 & 98.2\% \\
6 & 64 to 128 & 2.90 & 0.16 & 17.43 & 97.9\% \\
7 & 128 to 256 & 3.27 & 0.21 & 17.85 & 98.4\% \\
8 & 256 to 512 & 3.40 & 0.22 & 19.74 & 97.9\% \\
9 & 512 to 1024 & 3.52 & 0.28 & 19.12 & 98.6\% \\
10 & 1024 to 2048 & 3.35 & 0.26 & 24.81 & 98.5\% \\
11 & 2048 to 4096 & 3.03 & 0.24 & 29.66 & 97.7\% \\
12 & 4096 to 8192 & 2.44 & 0.10 & 16.89 & 95.7\% \\
13 & 8192 to 16384 & 2.18 & 0.10 & 15.15 & 85.5\% \\
14 & 16384 to 32768 & 1.99 & 0.32 & 5.70 & 82.4\% \\
15 & 32768 to 65536 & 1.98 & 0.23 & 9.51 & 87.0\% \\
16 & 65536 to 131072 & 1.73 & 0.30 & 4.59 & 79.3\% \\
17 & 131072 to 262144 & 1.33 & 0.51 & 3.31 & 58.0\% \\
$\ge 18$ & $\ge$ 262144 & $\approx$ 1.00 & 0.28 & 2.39 & 41 to 83\% \\
\bottomrule
\end{tabular}
\end{table}

\FloatBarrier

\subsection{CAFS vs.\ vqsort}

Bins 5 to 15 give a lead with bin means between 1.27 and 2.34 and win rates between 57\% and 93\%. The maximum, 2.34, is in bin 9 ($K = 512$ to 1024) — the same point as in the IPS4o pairing — covering $K$ between 32 and roughly 65000 (5 to 16 bits of entropy). CAFS wins here because the hash scheme runs at about one SIMD operation per element, while vqsort traverses $\log_2 N$ partition levels with a YMM sorting network at each level.

Bins 1, 3, and 4 are an anomaly: the bin mean dips to between 0.84 and 0.99 and the win rate falls to 19\% to 40\%, even though we would expect a comfortable lead at small $K$. Our best explanation is grid composition: most points in these bins are at small $N$, where the 100-microsecond sampling pre-pass is a substantial fraction of total run time and vqsort wins because it has no pre-pass. Bin 2 ($K = 4$ to 8) does not dip in the same way, which is consistent with that reading, because most bin-2 points fall into the tiny-count branch and bypass the Chao1 estimator. The cause has not been pinned down in our measurements, and this remains the weakest part of our characterisation of CAFS against vqsort at very low $K$.

From bin 16 the bin mean drops to 1.21, then to 1.01 at bin 17, and from bin 18 ($K \ge 262144$) it goes below 1, eventually reaching 0.58 to 0.63 in bins 21 to 24 with a zero win rate. At $K \approx N$ vqsort beats CAFS by a factor of 1.6 to 1.7 because its hot loop runs at 0.5 to 1 cycle per element through a YMM sorting network independent of $K$~\cite{wassenberg2022}, leaving CAFS with no remaining advantage.

The last dominance point at $N > 10^6$ corresponds to $K \approx 6.7 \cdot 10^5$ ($H \approx 19.36$ bits), past which the safety guard fires on a fraction of cases and keeps the lag near 1.5 to 2 times. Within the lead band the speedup is essentially flat in $N$ (Pearson $r = 0.072$ for $\log N$ vs.\ speedup). vqsort is therefore the only one of the five baselines that strictly overtakes CAFS in the high-$K$ regime: CAFS beats vqsort for $K \le 6.7 \cdot 10^5$, and above that point vqsort wins by a factor of 1.5 to 1.7.

\begin{figure}[H]
\centering
\includegraphics[width=0.82\textwidth]{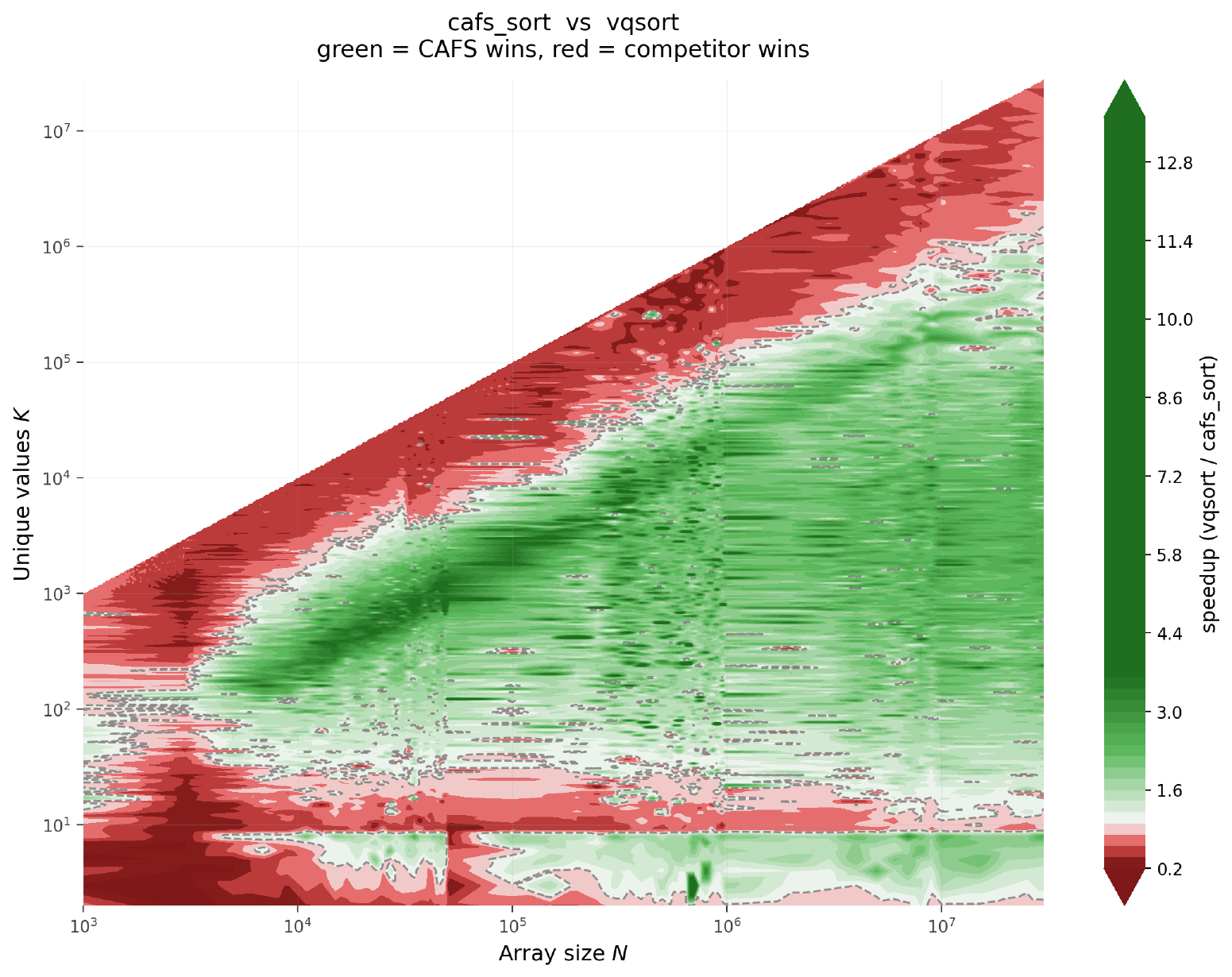}
\caption{Heatmap of CAFS speedup over vqsort: transition through parity at $H \approx 19.4$ bits.}
\label{fig:vs_vqsort}
\end{figure}

\begin{table}[H]
\centering
\caption{Speedup of CAFS over vqsort, by entropy bin.}
\label{tab:vs_vqsort}
\begin{tabular}{c c r r r r}
\toprule
$H_{\text{bin}}$ & $K$ range & avg & min & max & win-rate \\
\midrule
1 & 2 to 4 & 0.89 & 0.09 & 6.54 & 25.9\% \\
2 & 4 to 8 & 1.32 & 0.14 & 3.42 & 79.7\% \\
3 & 8 to 16 & 0.84 & 0.13 & 3.08 & 19.0\% \\
4 & 16 to 32 & 0.99 & 0.15 & 3.19 & 40.2\% \\
5 & 32 to 64 & 1.27 & 0.15 & 3.94 & 91.3\% \\
6 & 64 to 128 & 1.54 & 0.11 & 10.25 & 91.3\% \\
7 & 128 to 256 & 1.78 & 0.13 & 9.22 & 93.7\% \\
8 & 256 to 512 & 2.06 & 0.12 & 8.75 & 93.4\% \\
9 & 512 to 1024 & 2.34 & 0.14 & 18.10 & 93.1\% \\
10 & 1024 to 2048 & 2.18 & 0.14 & 10.53 & 91.1\% \\
11 & 2048 to 4096 & 1.91 & 0.18 & 16.34 & 83.0\% \\
12 & 4096 to 8192 & 1.64 & 0.06 & 14.77 & 59.9\% \\
13 & 8192 to 16384 & 1.59 & 0.05 & 14.24 & 57.6\% \\
14 & 16384 to 32768 & 1.45 & 0.18 & 9.42 & 60.6\% \\
15 & 32768 to 65536 & 1.39 & 0.12 & 5.37 & 65.4\% \\
16 & 65536 to 131072 & 1.21 & 0.16 & 4.18 & 49.9\% \\
17 & 131072 to 262144 & 1.01 & 0.26 & 3.48 & 39.1\% \\
18 & 262144 to 524288 & 0.77 & 0.22 & 2.06 & 28.9\% \\
19 & 524288 to 1048576 & 0.73 & 0.19 & 1.85 & 21.7\% \\
$\ge 20$ & $\ge$ 1048576 & 0.58 to 0.68 & 0.17 & 1.42 & 0 to 7.7\% \\
\bottomrule
\end{tabular}
\end{table}

\FloatBarrier

\subsection{CAFS vs.\ ska\_sort}

Against ska\_sort the curve has a different shape. The speedup in bins 1 and 2 reaches 17 times at a 100\% win rate, and in bins 3 to 11 the bin mean stays between 4.5 and 8 with win rates above 94\%, because ska\_sort runs eight byte passes over a 64-bit key whatever $K$ may be, while CAFS at small $K$ runs one SIMD compare per element. The runtime gap is the entropy gap of Section~\ref{sec:problem} written in microseconds.

The lead then contracts as $K$ grows: by bin 12 the bin mean is down to 3.2, by bin 16 to 1.7, and the last dominance point at $N > 10^6$ corresponds to $K \approx 8.14 \cdot 10^5$ ($H \approx 19.63$ bits). Past that boundary ska\_sort takes the lead, and in bins 22 to 24 ($K \in [4 \cdot 10^6, 3 \cdot 10^7]$) the bin mean is between 0.66 and 0.72 at a 0\% win rate. The root cause at high $K$ is memory pressure rather than instruction count: the CAFS hash table grows to tens of megabytes and loses L3 residency, so the hot loop becomes memory-bound, while ska\_sort works only on the input array and a small histogram, keeping its working set compact even when $K \approx N$.

Inside the dominance zone the lead is independent of $N$ ($r = 0.127$ between $\log N$ and speedup, near zero). The descent of the bin mean against ska\_sort is monotonic from bin~1 through bin~18, in contrast to the non-monotonic dip near small $K$ that appears against vqsort. The crossover at $K \approx 8.14 \cdot 10^5$ comes later than the vqsort crossover at $K \approx 6.7 \cdot 10^5$, so any deployment that may push $K$ beyond $10^6$ should fall back to vqsort, not ska\_sort, at the boundary.

\begin{figure}[H]
\centering
\includegraphics[width=0.82\textwidth]{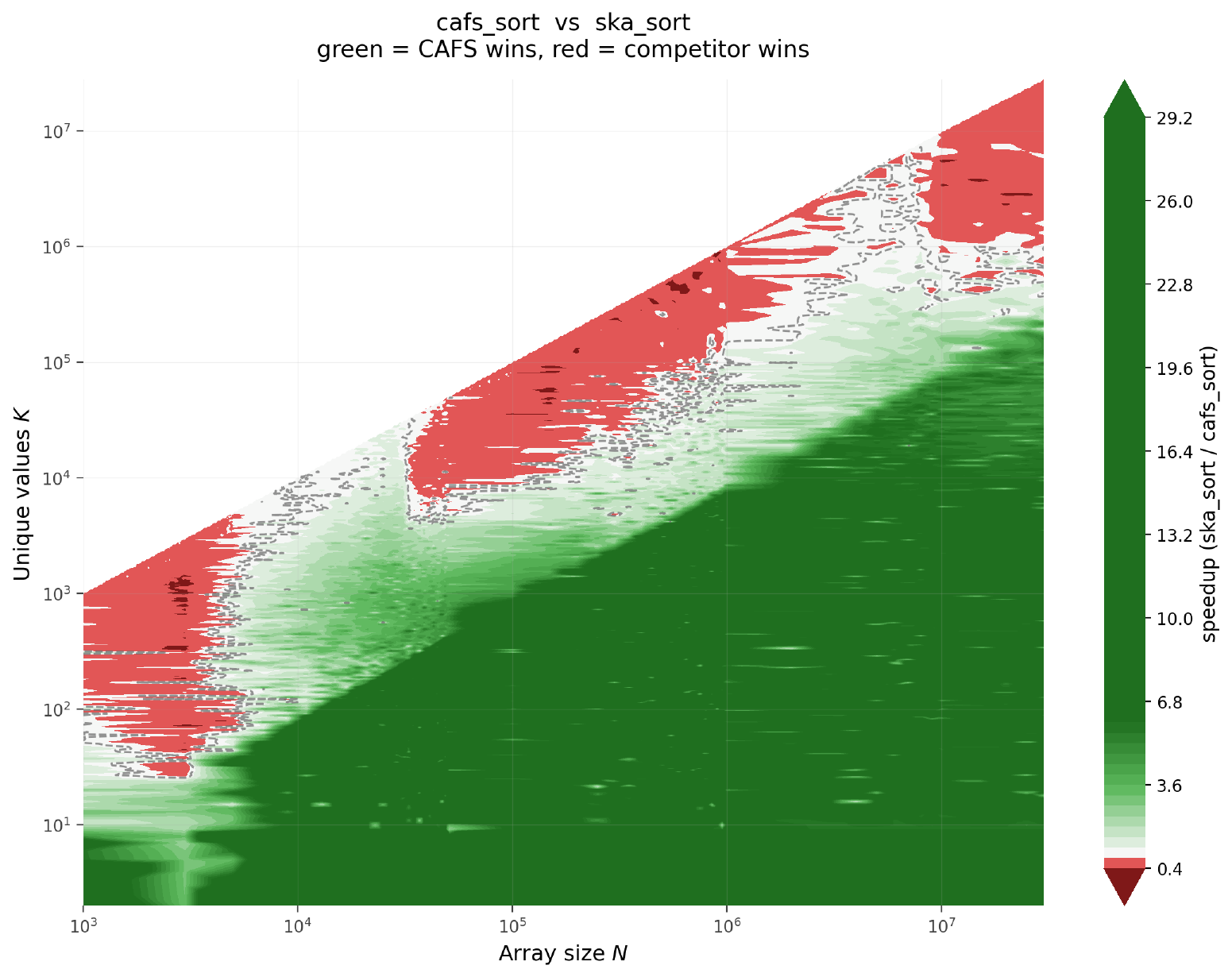}
\caption{Heatmap of CAFS speedup over ska\_sort across the $(N, K)$ grid.}
\label{fig:vs_ska}
\end{figure}

\begin{table}[H]
\centering
\caption{Speedup of CAFS over ska\_sort, by entropy bin.}
\label{tab:vs_ska}
\begin{tabular}{c c r r r r}
\toprule
$H_{\text{bin}}$ & $K$ range & avg & min & max & win-rate \\
\midrule
1 & 2 to 4 & 17.65 & 3.07 & 26.26 & 100.0\% \\
2 & 4 to 8 & 16.96 & 2.38 & 29.33 & 100.0\% \\
3 & 8 to 16 & 7.87 & 0.93 & 26.27 & 99.8\% \\
4 & 16 to 32 & 7.03 & 0.39 & 12.26 & 99.4\% \\
5 & 32 to 64 & 7.60 & 0.21 & 13.13 & 97.8\% \\
6 & 64 to 128 & 7.64 & 0.10 & 19.64 & 95.1\% \\
7 & 128 to 256 & 7.43 & 0.23 & 27.86 & 95.6\% \\
8 & 256 to 512 & 6.19 & 0.24 & 15.15 & 95.2\% \\
9 & 512 to 1024 & 5.87 & 0.17 & 14.41 & 95.2\% \\
10 & 1024 to 2048 & 5.38 & 0.20 & 18.39 & 95.9\% \\
11 & 2048 to 4096 & 4.49 & 0.27 & 13.67 & 94.0\% \\
12 & 4096 to 8192 & 3.17 & 0.13 & 12.79 & 78.9\% \\
13 & 8192 to 16384 & 2.45 & 0.18 & 9.21 & 74.6\% \\
14 & 16384 to 32768 & 2.19 & 0.20 & 7.33 & 63.0\% \\
15 & 32768 to 65536 & 2.07 & 0.21 & 7.61 & 56.1\% \\
16 & 65536 to 131072 & 1.69 & 0.21 & 6.87 & 50.8\% \\
17 & 131072 to 262144 & 1.26 & 0.27 & 5.34 & 41.2\% \\
18 & 262144 to 524288 & 0.89 & 0.29 & 2.22 & 37.2\% \\
$\ge 19$ & $\ge$ 524288 & $\le$ 0.87 & 0.23 & 2.09 & $\le$ 35.7\% \\
\bottomrule
\end{tabular}
\end{table}

\FloatBarrier

\subsection{CAFS vs.\ std::sort}

CAFS beats std::sort in every entropy bin, including the $K \approx N$ regime. Across the band $H_{\text{bin}} \in [1, 16]$ the bin-mean speedup is between 7 and 17 at win rates above 97\%. Even in the high-entropy zone (bins 22 to 24) the win rate is 100\% with a 3.5 times speedup, driven by the CAFS fallback to pdqsort, since std::sort (introsort) loses to pdqsort across the whole grid and CAFS routes to pdqsort in the high-$K$ branch. The last dominance point at $N > 10^6$ corresponds to $K \approx 1.72 \cdot 10^7$ ($H \approx 24$ bits), and the lead grows with $N$ ($r = 0.351$ between $\log N$ and speedup in the dominance zone) because introsort runs data-dependent branches whose mispredictions accumulate with $N$ while the CAFS hot loop has no comparable branch. There is no point on the grid at which CAFS loses to std::sort.

\begin{figure}[H]
\centering
\includegraphics[width=0.82\textwidth]{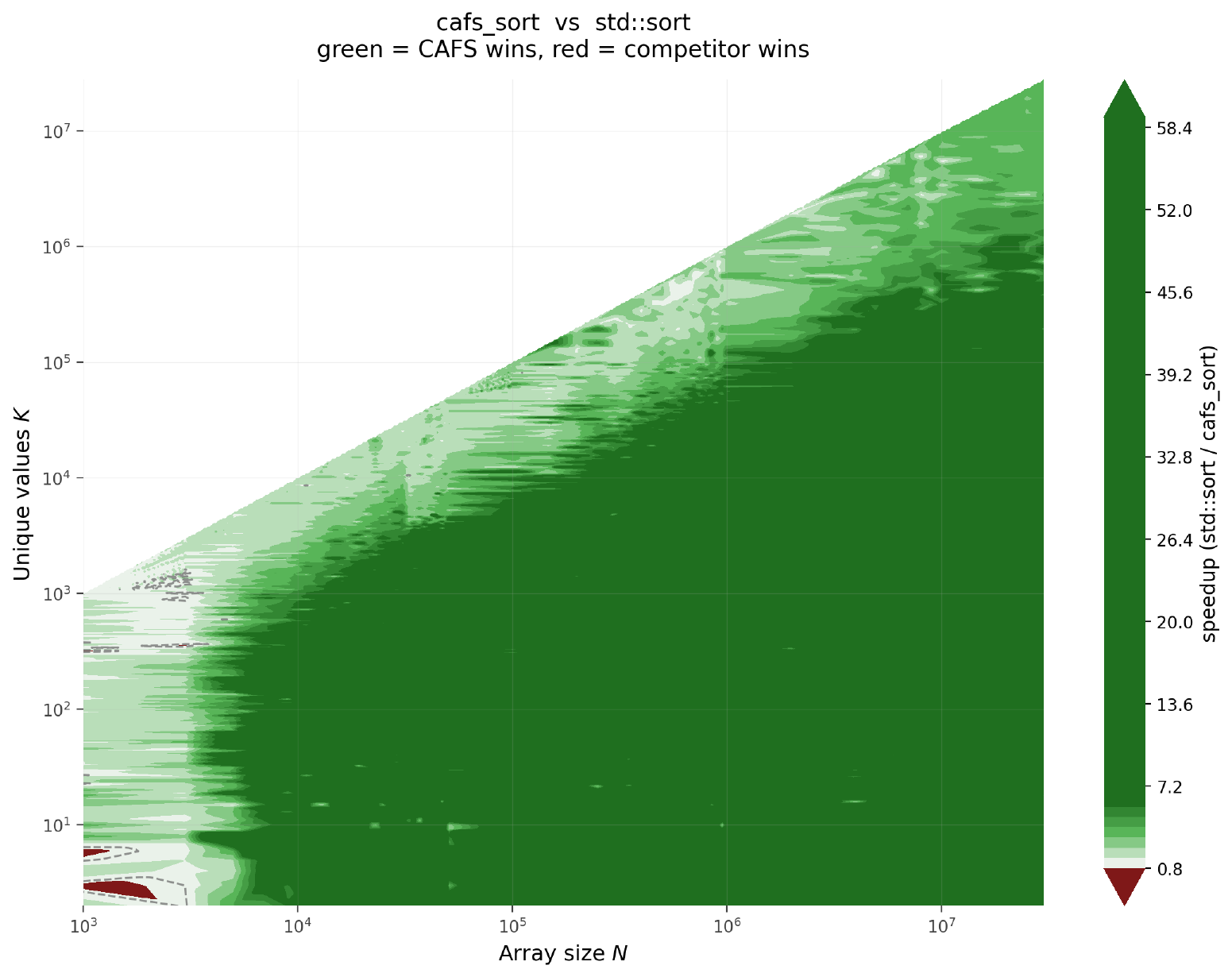}
\caption{Heatmap of CAFS speedup over std::sort across the $(N, K)$ grid.}
\label{fig:vs_stdsort}
\end{figure}

\begin{table}[H]
\centering
\caption{Speedup of CAFS over std::sort, by entropy bin.}
\label{tab:vs_stdsort}
\begin{tabular}{c c r r r r}
\toprule
$H_{\text{bin}}$ & $K$ range & avg & min & max & win-rate \\
\midrule
1 & 2 to 4 & 13.09 & 0.48 & 23.00 & 97.4\% \\
2 & 4 to 8 & 17.10 & 0.58 & 58.96 & 99.1\% \\
3 & 8 to 16 & 9.67 & 0.91 & 60.62 & 99.8\% \\
4 & 16 to 32 & 10.02 & 0.87 & 47.67 & 99.8\% \\
5 & 32 to 64 & 12.21 & 0.98 & 36.30 & 99.9\% \\
6 & 64 to 128 & 13.52 & 0.65 & 67.39 & 99.9\% \\
7 & 128 to 256 & 14.51 & 0.56 & 90.38 & 99.9\% \\
8 & 256 to 512 & 14.76 & 0.41 & 44.22 & 99.7\% \\
9 & 512 to 1024 & 14.74 & 0.50 & 60.89 & 99.7\% \\
10 & 1024 to 2048 & 14.08 & 0.61 & 66.57 & 99.6\% \\
11 & 2048 to 4096 & 12.74 & 0.76 & 60.93 & 99.9\% \\
12 & 4096 to 8192 & 10.71 & 0.24 & 57.79 & 99.9\% \\
13 & 8192 to 16384 & 9.46 & 0.22 & 44.23 & 99.9\% \\
14 & 16384 to 32768 & 8.56 & 0.74 & 32.32 & 99.6\% \\
15 & 32768 to 65536 & 8.34 & 0.54 & 23.88 & 100.0\% \\
16 & 65536 to 131072 & 7.15 & 0.85 & 21.36 & 99.9\% \\
17 & 131072 to 262144 & 5.30 & 1.18 & 20.20 & 100.0\% \\
18 & 262144 to 524288 & 3.93 & 1.12 & 11.29 & 100.0\% \\
19 & 524288 to 1048576 & 3.70 & 0.95 & 9.14 & 99.2\% \\
$\ge 20$ & $\ge$ 1048576 & $\ge$ 3.20 & 0.92 & 6.69 & $\ge$ 94.8\% \\
\bottomrule
\end{tabular}
\end{table}

\FloatBarrier

\subsection{Summary applicability map}

Aggregating across the five baselines yields the applicability map of CAFS in $(N, K)$ coordinates.

\begin{figure}[!htb]
\centering
\includegraphics[width=0.88\textwidth]{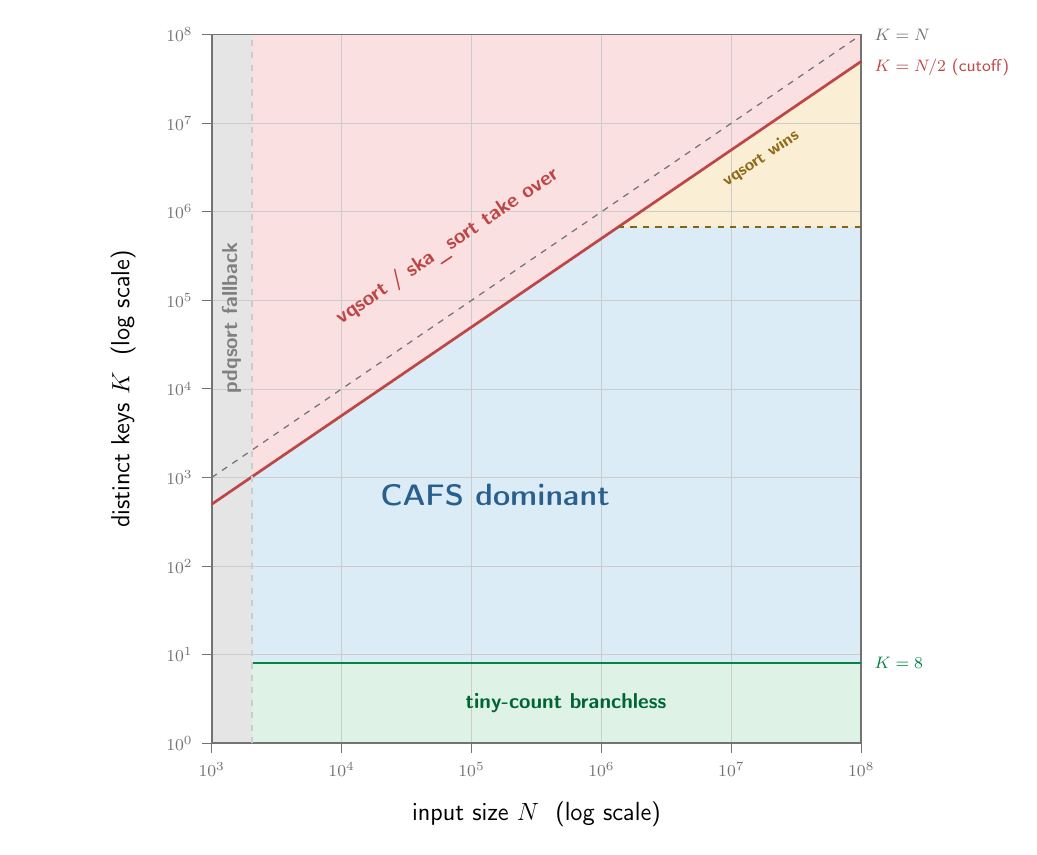}
\caption{Applicability map of CAFS in $(N, K)$ coordinates.}
\label{fig:applicability}
\end{figure}

\begin{table}[!htb]
\centering
\caption{Applicability zones of CAFS.}
\label{tab:zones}
\begin{tabular}{p{2.7cm} p{2.5cm} p{9.0cm}}
\toprule
Zone & Parameters & CAFS behavior \\
\midrule
tiny-count & $K \le 8$, $N \ge 2048$ & beats all five baselines: 13 to 21 times vs std::sort, 2.5 to 5 times vs pdqsort, 17 to 19 times vs ska\_sort, 3 to 4 times vs IPS4o, 1 to 2 times vs vqsort \\
main CAFS & $8 < K \le 10^5$ & beats all five baselines: 1.7 to 3 times vs pdqsort, 4 to 8 times vs ska\_sort, 8 to 15 times vs std::sort, 1.7 to 3.5 times vs IPS4o, 1.2 to 2.3 times vs vqsort \\
transitional & $10^5 < K \le N/2$ & bin-mean speedup vs pdqsort, ska\_sort, IPS4o, vqsort drops toward 1; vs std::sort stays above 3 \\
high-entropy & $K \approx N$ & loses to vqsort (0.58 to 0.63 times) and ska\_sort (about 0.66 times); parity with IPS4o and pdqsort; lead over std::sort \\
pathological & $N < 2048$ & triggers fallback; total time equals pdqsort plus a 100-microsecond sampling overhead \\
\bottomrule
\end{tabular}
\end{table}

\FloatBarrier

The boundaries are: $K \approx 1.3 \cdot 10^5$ ($H \approx 17$ bits) against pdqsort and std::sort, $K \approx 8.14 \cdot 10^5$ ($H \approx 19.63$ bits) against ska\_sort, $K \approx 6.7 \cdot 10^5$ ($H \approx 19.36$ bits) against vqsort, and only $K \approx 1.72 \cdot 10^7$ (parity at $K = N$) against IPS4o. Of the five baselines, only vqsort strictly overtakes CAFS once its crossover is passed: against pdqsort and std::sort the algorithm falls back to pdqsort and matches it, against ska\_sort the lag is modest, and against IPS4o the curves only converge at $K = N$. The binding constraint on the operational range of CAFS is therefore the vqsort threshold at $K \approx 6.7 \cdot 10^5$.

Coming back to the three questions of Section~\ref{sec:eval}.1. The bin-mean speedup at $K \le 10^3$ is comfortably above 2$\times$ against std::sort, pdqsort, ska\_sort, and IPS4o, and between 1 and 2.3$\times$ against vqsort (with a few low-$K$ bins below 1 that we do not fully explain; see Section~\ref{sec:eval}.5). The per-baseline crossovers are the four numbers above. The lead at fixed $K$ is roughly stable in $N$ within the dominance zone: the Pearson correlations between $\log N$ and speedup are all small in magnitude ($|r| < 0.4$ at $\approx 24000$ measurements per pair) and the signs are consistent with the relative cost models, but with $|r|$ this small they do not by themselves rule out a stationary lead. The spill safety guard (Section~\ref{sec:cafs}) keeps the worst case observed in the grid under a 5$\times$ slowdown.

\section{Discussion and Limitations}\label{sec:disc}

\subsection{Operating-point guidance}

CAFS is designed for a specific input class and degrades to pdqsort outside it. For deployment in OLAP engines the relevant operating point is the post-group-by sort, where the cardinality of the key column is bounded by the number of distinct group keys. That number is typically below $10^5$, which falls inside the main CAFS zone for every baseline considered. Beyond OLAP, the main CAFS zone also covers categorical column sorting and foreign-key sorting against a small dimension table.

We recommend CAFS whenever an upper bound on $K$ is available and falls below $10^5$, vqsort whenever $K$ may exceed $6.7 \cdot 10^5$, and the per-baseline crossover for cases between the two thresholds. The CAFS dispatcher implements a runtime version of the same rule through the $\hat K \cdot 2 > N$ guard, although a deployment that knows the schema can short-circuit the pre-pass entirely.

\subsection{Why CAFS wins on $K \ll N$}

Comparison and radix sorts pay a per-element cost independent of the input entropy ($\log_2 N$ comparisons for the former, $\lceil B/b \rceil$ digit visits for the latter), whereas a hash-count sort pays one hash and one bucket compare per element when the bucket is sized correctly. With the bucket on one cache line and the compare folded into a single SIMD instruction, this drops to about 4 to 5 cycles per element, and in the main zone CAFS runs at that rate. The other stages (sampling, dispatcher, reconstruction) cost $O(s)$ or $O(K)$ and become negligible at large $N$.

\subsection{Why CAFS loses at $K \approx N$}

At $K \approx N$ the algorithm has nothing left to exploit. The hash table grows linearly with $\hat K$ and now occupies tens of megabytes, so it falls out of L3 into DRAM and the hot loop becomes memory-bound. The trailing pair sort, which is negligible at $K \ll N$, now runs on $K \approx N$ items at a cost of $O(N \log N)$, equal to a full comparison sort but with extra setup. The dispatcher condition $\hat K \cdot 2 > N$ catches most such cases at the pre-pass and reroutes them to pdqsort directly, and the safety guard catches the rest after the hot loop runs. The net cost is pdqsort time plus the 100 microseconds of sampling overhead that we never recover.

\subsection{Limitations}

Several limitations apply to the current evaluation. All measurements come from one hardware platform, an Intel Core i5-12400F, so behaviour on parts with different cache geometry (Zen, Skylake-X, server-class Xeon, mobile Alder Lake E-cores) remains untested. The L1d cache on our test platform is 48 KB, and on parts with a 32 KB L1d the threshold at which the table leaves L1d shifts down by 0.5 bits in $H$, which would compress the main CAFS zone slightly. The AVX-512 path of vqsort is not in our measurements, even though on supporting CPUs it is roughly 1.5 times faster than the AVX2 path and would push the vqsort crossover toward smaller $K$.

The benchmark is also restricted to single-threaded runs, so we do not compare against the parallel modes of pdqsort or IPS4o, and a parallel CAFS with per-thread hash tables and a final merge is plausible but not built. The grid uses a single input distribution, a uniform random palette of 64-bit values, while real workloads often have sorted-by-group, runs, or almost-sorted profiles where RLE injection would push CAFS further ahead. Because the multiplicative hash is deterministic and public, an adversary can construct inputs that drive collisions; no such inputs appear in our grid. Finally, the reported correlations between $\log N$ and speedup all have $|r| < 0.4$ at sample sizes near 24000 per pairing, so $N$ is not the dominant factor inside the dominance zone.

\subsection{Future directions}

The two short-term next steps are an AVX-512 bucket with eight 64-bit lanes (which should push the hot-loop throughput higher and move the vqsort crossover out) and a randomized hash seed (which closes the deterministic-collision attack noted above). The harder open items are a parallel CAFS with per-thread hash tables and a final merge, a floating-point variant via bit-cast, and a learned cardinality estimator that could replace Chao1 on workloads where the singletons-doubletons signal is weak. Two further items are integrating CAFS into a real query engine (DuckDB, ClickHouse, or MonetDB as the post-group-by sort) and extending the input grid past the uniform random palette to sorted-by-group, runs, and almost-sorted profiles. The grid extension is the most directly testable because RLE injection inside CAFS targets exactly those input shapes.

A v2 of this paper is planned for autumn 2026. The platform moves to an AMD Ryzen 5 9600X with 32\,GB DDR5-6000 on a bare Linux setup, wall-clock timing through \texttt{std::chrono} is replaced by \texttt{perf stat}, and the AVX2 hot loop is rewritten in AVX-512.

IPS4o is dropped from the baseline set in v2 to keep every competitor single-header, and a plain hash-count sort over \texttt{std::unordered\_map} is added. Informal runs already show CAFS beats it by orders of magnitude at every $K$ we tried; v2 will report the full grid against it in Evaluation.

Suggestions for additional baselines to include in v2, or proposals for collaborative research, are welcome.

\section{Conclusion}\label{sec:concl}

CAFS targets a specific corner of integer sorting: inputs where $K$ is much smaller than $N$. The algorithm pays a 100-microsecond Chao1 pre-pass to detect that regime, dispatches accordingly, and when it lands inside the regime runs the count-emit loop at about one SIMD operation per element. Two costs follow from the $K$-dependent working set. A spill safety guard catches inputs where the dispatcher misestimated $K$. And on inputs with $K$ close to $N$ the hash table eventually exceeds L3 cache, which is what gives vqsort the crossover at $K \approx 6.7 \cdot 10^5$ on our platform.

On the 58-by-dense-$K$ grid and inside the $K \ll N$ band, CAFS runs 1.7 to 3.1$\times$ faster than pdqsort, 1.7 to 3.5$\times$ faster than IPS4o, 4 to 17$\times$ faster than ska\_sort, 8 to 17$\times$ faster than std::sort, and 1.2 to 2.3$\times$ faster than vqsort, with the operational range upper-bounded by the vqsort crossover at $K \approx 6.7 \cdot 10^5$. Two questions remain open: whether an AVX-512 bucket and a randomized hash seed can move that crossover above $K = 10^7$, and how the picture changes on skewed inputs and on sorted-by-group profiles. The implementation and the raw measurements are available at \url{https://github.com/kexibq-official/cafs-lib}.

\bibliography{ms}

\end{document}